\documentclass[twocolumn]{aastex61}
\usepackage{amsmath}
\usepackage{amssymb}
\usepackage{bm}
\usepackage{courier}
\usepackage{color}
\usepackage{comment}
\usepackage{mathtools}
\usepackage{hyperref}

\begin{document}

%@arxiver{result_paper_wd1_rc.png}

\title{The Optical/Near-Infrared Extinction Law in Highly Reddened Regions}
\author{Matthew W. Hosek Jr.}
\affiliation{Institute for Astronomy, University of Hawaii, 2680 Woodlawn Drive, Honolulu, HI 96822, USA}
\correspondingauthor{Matthew W. Hosek Jr.}
\email{mwhosek@ifa.hawaii.edu}

\author{Jessica R. Lu}
\affiliation{Department of Astronomy, 501 Campbell Hall, University of California, Berkeley, CA, 94720}
\affiliation{Institute for Astronomy, University of Hawaii, 2680 Woodlawn Drive, Honolulu, HI 96822, USA}

\author{Jay Anderson}
\affiliation{Space Telescope Science Institute, 3700 San Martin Drive, Baltimore, MD 21218, USA}

\author{Tuan Do}
\affiliation{UCLA Department of Physics and Astronomy, Los Angeles, CA 90095}

\author{Edward F. Schlafly}
\affiliation{Lawrence Berkeley National Laboratory, One Cyclotron Road, Berkeley, CA 94720, USA}

\author{Andrea M. Ghez}
\affiliation{UCLA Department of Physics and Astronomy, Los Angeles, CA 90095}

\author{William I. Clarkson}
\affiliation{Department of Natural Sciences, University of Michigan-Dearborn, 4901 Evergreen Road, Dearborn, MI 48128}

\author{Mark R. Morris}
\affiliation{UCLA Department of Physics and Astronomy, Los Angeles, CA 90095}

\author{Saundra M. Albers}
\affiliation{Department of Astronomy, 501 Campbell Hall, University of California, Berkeley, CA, 94720}

\begin{abstract}

A precise extinction law is a critical input when interpreting observations of highly reddened sources such as young star clusters and the Galactic Center (GC). We use Hubble Space Telescope observations of a region of moderate extinction and a region of high extinction to measure the optical and near-infrared extinction law (0.8 $\mu$m -- 2.2 $\mu$m). The moderate extinction region is the young massive cluster Westerlund 1 (Wd1; A$_{Ks} \sim$ 0.6 mag), where 453 proper motion-selected main-sequence stars are used to measure the shape of the extinction law. To quantify the shape we define the parameter $\mathcal{S}_{1/\lambda}$, which behaves similarly to a color excess ratio but is continuous as a function of wavelength. The high extinction region is the GC (A$_{Ks} \sim$ 2.5 mag), where 819 red clump stars are used to determine the normalization of the law. The best-fit extinction law is able to reproduce the Wd1 main sequence colors, which previous laws misestimate by 10\%~--~30\%. The law is inconsistent with a single power law, even when only the near-infrared filters are considered, and has A$_{F125W}$~/~A$_{Ks}$ and A$_{F814W}$~/~A$_{Ks}$ values that are 18\% and 24\% larger than the commonly used \citet{Nishiyama:2009fc} law, respectively. Using the law we recalculate the Wd1 distance to be 3896 $\pm$ 328 pc from published observations of eclipsing binary W13. This new extinction law should be used for highly reddened populations in the Milky Way, such as the Quintuplet cluster and Young Nuclear Cluster. A python code is provided to generate the law for future use.

\end{abstract}

\section{Introduction}
\label{sec:intro}
Understanding the optical through near-infrared (OIR; I--K band; 0.8 $\mu$m -- 2.2 $\mu$m) extinction law is critically important for studying objects beyond the solar neighborhood. For example, measurements of the stellar initial mass function in extinguished clusters \citep[e.g.][]{Habibi:2013th}, studies of stellar populations at the Galactic Center \citep[e.g.][]{Feldmeier-Krause:2015wk}, and mapping the Milky Way's structure \citep[e.g.][]{Bovy:2016mg} all depend on precise and accurate knowledge of the extinction law. In addition, the shape of the extinction law depends on the underlying dust grain properties along the line of sight, and so measuring the extinction law provides insight into grain characteristics across different interstellar environments \citep[e.g.][]{Draine:2003ij, Voshchinnikov:2017lr}. Commonly used extinction laws such as \citet{Cardelli:1989qf} and \citet{Fitzpatrick:1999fj} describe the OIR extinction as a power law ($A_{\lambda} \propto \lambda^{-\beta}$) with $\beta$ = 1.6, derived from photometry of a small sample of stars along different sightlines. Additional studies at that time obtained similar results with $\beta$ values of $\sim$1.7 -- 1.8 \citep[e.g.][]{Draine:1989th, Martin:1990sp}, and so this description of the OIR extinction law has been widely adopted.

More recent studies have used large-scale photometric surveys to refine the measurement of the OIR extinction law. A power law has been found to be a good fit in the NIR regime (J--K; 1.25 $\mu$m -- 2.2 $\mu$m), though with a generally steeper exponent than was found in earlier work. Studies such as \citet{2005ApJ...619..931I}, \citet{Messineo:2005uk}, and \citet{Nishiyama:2009fc} use Two-Micron All Sky Survey (2MASS) JHK photometry of red giant stars (often red clump stars) to measure extinction, reporting $\beta$ = 1.7, 1.9, and 2.0 respectively. The steeper slopes of \citet{Messineo:2005uk} and \citet{Nishiyama:2009fc} were derived for fields toward the Galactic Bulge, while the shallower slope of \citet{2005ApJ...619..931I} was found for fields at Galactic longitudes of $\ell$ = 42$^{\circ}$ and $\ell$ = 284$^{\circ}$, hinting at a variation in the law with Galactic longitude. However, spectroscopic studies of red clump stars in the APOGEE \citep{Wang:2014xr} and \emph{Gaia}-ESO surveys \citep{Schultheis:2015di} show no evidence of NIR law variability as a function of either total extinction or angle relative to the Galactic center. These studies find slopes of 1.95 and 2.12, respectively, supporting a steeper NIR extinction law.

A more complicated function is required for the extinction law when observations shortward of J-band are included. The \citet{Cardelli:1989qf} law has a single free parameter, R$_V$ (the ratio of absolute to selective extinction in V-band), which begins to significantly impact the steepness of the law shortward of I-band. \citet{Fitzpatrick:2009ys} adopt a two-parameter model  model that behaves as a power law whose exponent increases with wavelength in order to reproduce Hubble Space Telescope (\emph{HST}) spectrophotometry (0.75 $\mu$m -- 1 $\mu$m) and ground-based JHK photometry for a sample of OB-type stars. A study of red clump (RC) stars in the OGLE-III and VVV surveys by \citet{Nataf:2016dd} similarly concluded that a law with multiple free parameters is required to reproduce the observed colors in VJHK (0.5 -- 2.14 $\mu$m), finding the \citet{Cardelli:1989qf} law to be inconsistent regardless of how R$_V$ is varied. However, \citet{Schlafly:2016cr} present a one-parameter law where variations in R$_V$ can explain Pan-STARRS, 2MASS, and WISE photometry (0.5 $\mu$m -- 4.5 $\mu$m) of APOGEE stars in the galactic disk, though this law is most similar to the \citet{Fitzpatrick:2009ys} law.

While able to utilize large samples across many lines of sight, a disadvantage of survey-based extinction studies is that they are limited by the photometric depth of the surveys used and are dominated by low-extinction stars. The Galactic Center (GC) presents an opportunity to measure the extinction law at high extinctions (A$_{Ks} \sim$ 2.5 mag), where small variations in the law can have a large effect on observations. Previous studies of the GC extinction law include \citet{Schodel:2010eq}, who use photometry of RC stars in the central parsec region to measure a power law slope of $\beta$ = 2.21 $\pm$ 0.24 between H- and K-band, and \citet{Fritz:2011cr}, who measure $\beta$ = 2.11 $\pm$ 0.06 from analysis of gaseous emission lines ($\lambda$ $\geq$ 1 $\mu$m) from the minispiral structure near Sgr A*. Most recently, \citet{2017arXiv170909094N} obtain $\beta$~=~2.31 $\pm$ 0.03 for JHK observations of RC stars in the wide-field (7.95' x 3.43') GALACTICNUCLEUS survey, finding no dependence on field location or total extinction. However, the trade-off for observing at such high reddening is the rapid loss of starlight shortward of J-band, and so the optical portion of the GC extinction law remains largely unexplored.

In this paper, we combine \emph{HST} observations of a region of moderate extinction with a region of high extinction in order to constrain the shape and normalization of the extinction law between 0.8 $\mu$m -- 2.2 $\mu$m. The moderate extinction region is Westerlund 1 (Wd1), a young massive cluster with an extinction of A$_{Ks} \sim$ 0.7 mags \citep[][hereafter D16]{Damineli:2016no} located in the Galactic plane ($\ell$ = -20.451$^{\circ}$, $b$ = -0.404$^{\circ}$). While these observations allow us to sensitively probe the shape of the extinction law through the I-band, the uncertainty in the cluster distance prevents a tight constraint on the normalization of the law. To overcome this, we incorporate NIR \emph{HST} observations of red clump (RC) stars found in the line-of-sight (LOS) toward the Arches cluster, a similar young massive cluster near the GC ($\ell$ = 0.121$^{\circ}$, $b$ = 0.0168$^{\circ}$). The average distance of these stars is much better constrained and thus the normalization of the law can be determined. We use a forward-modeling Bayesian technique to simultaneously fit the extinction law and global properties of both populations without assuming a functional form for the law.

\section{Observations and Measurements}
\label{sec:obs}
\begin{figure*}
\begin{center}
\includegraphics[scale=0.5]{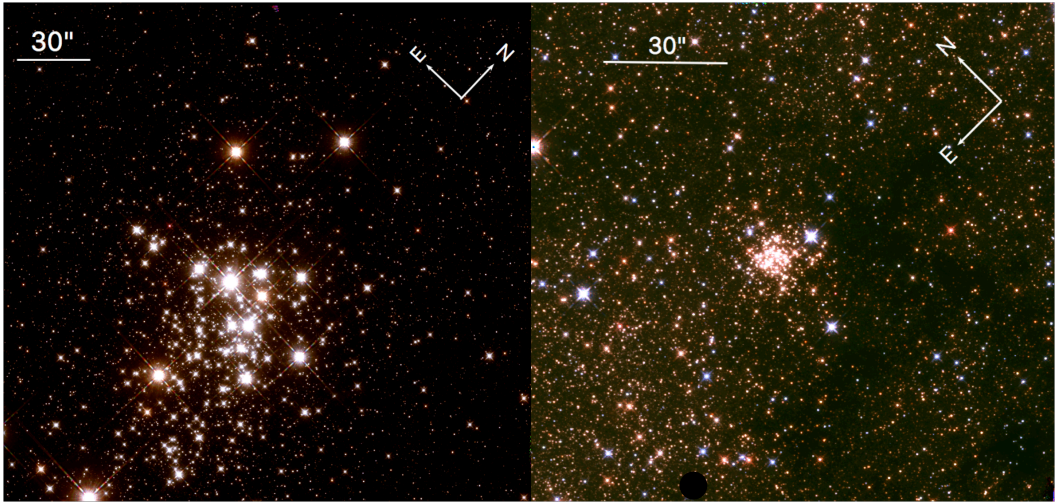}
\end{center}
\caption{
\emph{HST} three-color images of Wd1 (left) and the Arches cluster field (right). The Wd1 image is a 2 x 2 mosaic created with the WFC3-IR camera, with F125W as blue, F139M as green, and F160W as red. The Arches image is a single WFC3-IR field with F127M as blue, F139M as green, and F153M as red. The F139M observations are not used in the extinction law analysis.
\label{fig:rgb}
}
\end{figure*}

\begin{figure}
\begin{center}
\includegraphics[scale=0.18]{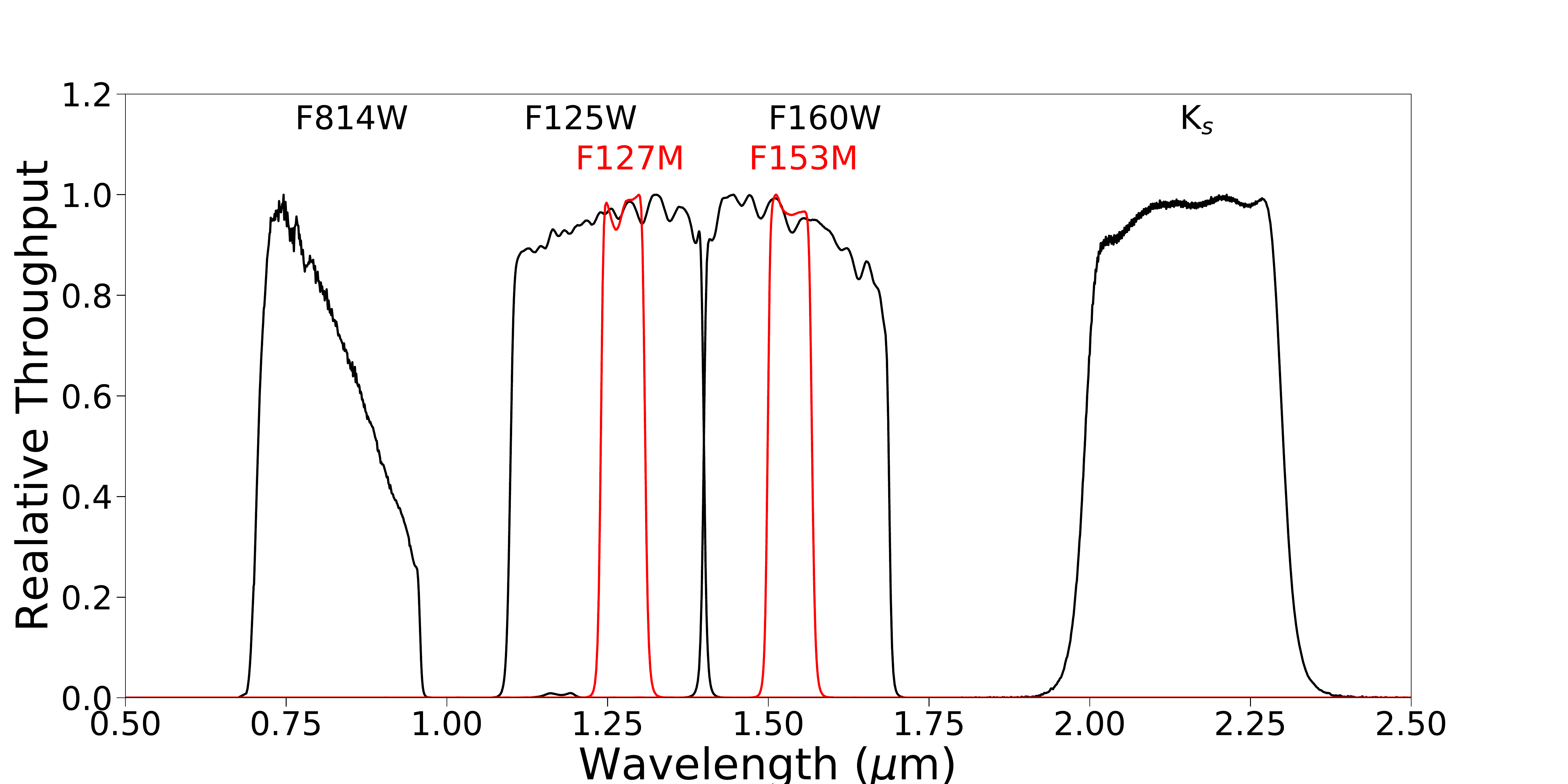}
\end{center}
\caption{
Filters used to constrain the Wd1 and Arches field RC star extinction law. The F125W, F127M, F153M, and F160W filters are from \emph{HST} WFC3-IR, the F814W filter from \emph{HST} ACS-WFC3, and the K$_{s}$ filter from the \emph{VISTA} VVV survey.
\label{fig:filters}
}
\end{figure}

We combine observations of Wd1 and RC stars in the Arches cluster field to measure the OIR extinction law (Figure \ref{fig:rgb}). For Wd1, we use \emph{HST} observations in the F814W, F125W, and F160W filters combined with VISTA K$_s$ observations obtained through the \emph{VISTA Variables in the Via Lactea} (\emph{VVV}) survey \citep{Minniti:2010lp}. For the RC stars, we use \emph{HST} observations obtained in the F127M and F153M filters (1.27 $\mu$m, and 1.53 $\mu$m, respectively). These filters provide photometry from 0.8 $\mu$m -- 2.2 $\mu$m with the throughputs shown in Figure \ref{fig:filters}. An overview of the observations is provided in Table \ref{tab:HST_obs}.

\subsection{HST Observations: Wd1}
\label{sec:Wd1HSTobs}
\emph{HST} observations of Wd1 were obtained over an 8 year period between 2005 - 2013. The earliest observations were made in 2005 with the Advanced Camera for Surveys Wide Field Camera (ACS-WFC, GO-10172, PI: R. De Grijs) in the F814W filter. The total exposure time was 2407 s, comprised of 3 slightly dithered images covering a 211" x 218" field of view (0".05 pix$^{-1}$). A second set of observations was obtained in 2010 with the infrared channel of the Wide Field Camera 3 (WFC3-IR) in the F125W and F160W filters (GO-11708, PI: M. Andersen)\footnote{Observations were also obtained in the F139M filter, but a comparison between the output photometry and stellar SED models indicated either a problem with the photometry (perhaps an incorrect zeropoint) or stellar models at these wavelengths (1.35~$\mu$m~--~1.41~$\mu$m). As a result, F139M is not considered in this analysis.}. The final observations were obtained in 2013 with WFC3-IR in the F160W filter to provide a third positional epoch (GO-13044, PI: J. R. Lu). These observations mimicked the 2010 F160W observations, though a different position angle was used due to new \emph{HST} guide star restrictions. Since the field of view for WFC3-IR is $\sim$130'' (0".12 pix$^{-1}$), a 2 x 2 mosaic was used to cover the entire 2005 ACS-WFC field of view. Each pointing had 7 images per filter, for total exposure times of 2443 s and 2093 s in F125W and F160W, respectively.

The \emph{HST} observations were reduced using the standard online HST data reduction pipeline and the resulting ``FLT'' images were downloaded from the HST archive on 2011 Dec 14. High-quality astrometric and photometric measurements were extracted from the individual FLT images using \emph{KS2}, an expansion of the software developed for the Globular Cluster Treasury Program \citep{Anderson:2008qy}. With this software, the measurements are combined to produce a single starlist for each filter. The sources are matched across epochs and positions transformed into a common astrometric reference frame where proper motions can be calculated. A detailed explanation of this process is given in H15.

\begin{figure*}
\begin{center}
\includegraphics[scale=0.7]{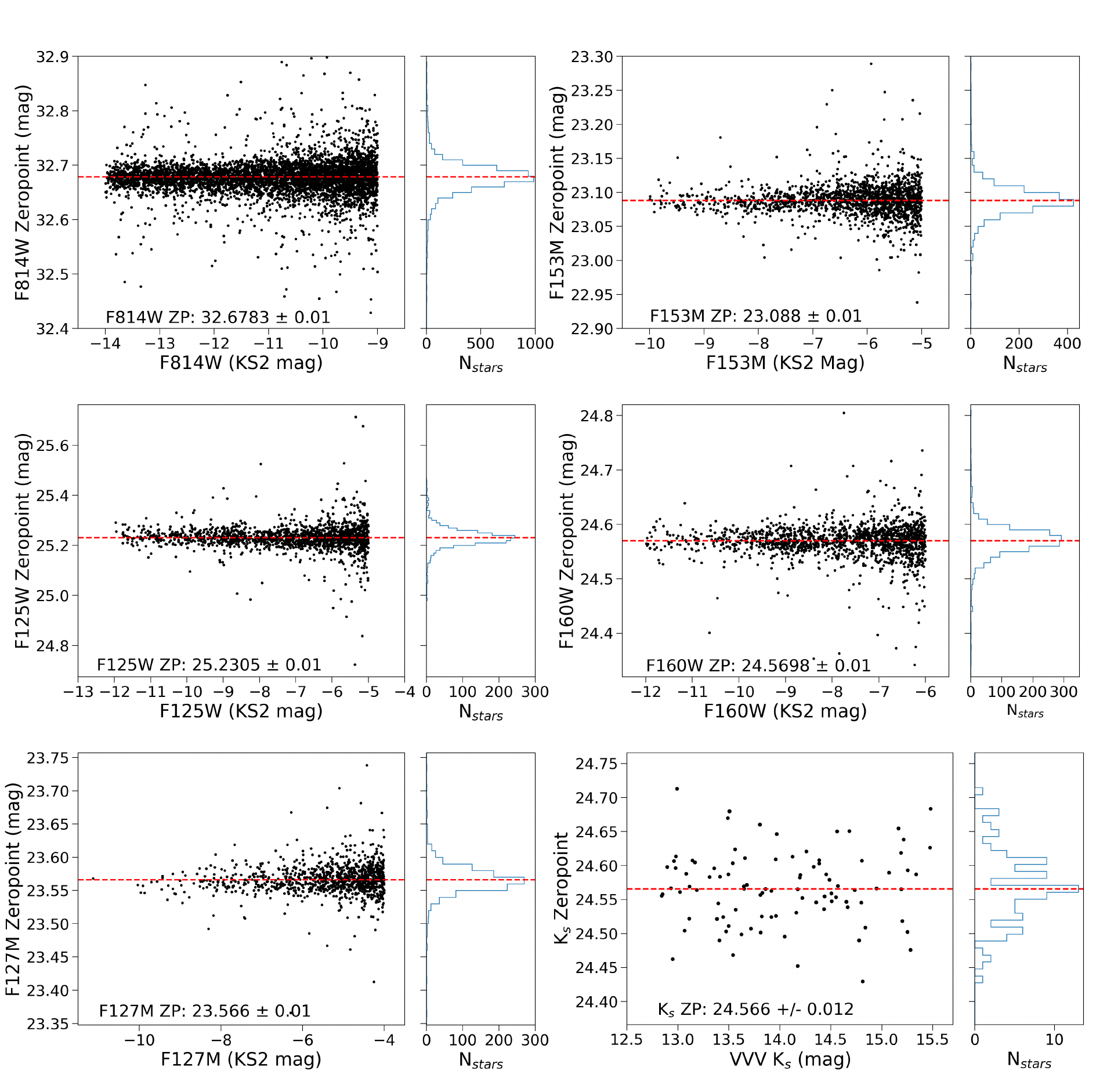}
\end{center}
\caption{The photometric zeropoints derived for the filters used in this study. The \emph{HST} filter zeropoints are for the \emph{KS2} photometry and the \emph{VISTA} K$_s$ zeropoint is for the \emph{AIROPA} photometry. The \emph{KS2} zeropoints have uncertainties of 0.01 mags and the \emph{AIROPA} zeropoint has a statistical uncertainty of 0.012 mags.
\label{fig:ZP}
}
\end{figure*}

To calculate the KS2 photometric zeropoints, we compared the KS2 starlists with calibrated photometry obtained using DOLPHOT, a version of HSTPHOT \citep{Dolphin:2000mw} with specific modules for ACS and WFC3-IR. We used Tiny Tim \citep{Krist:2011ev} point spread functions and the general procedure and recommendations (e.g., DOLPHOT input parameters) outlined in \citet{Williams:2014rp}. The DOLPHOT output was culled to eliminate spurious and overly crowded sources using sharpness, crowding, and SNR cuts of $<$1, $<$0.55, and $>$5, respectively. The remaining sources were cross-matched with the KS2 starlists, and the KS2 zeropoints calculated from the average difference between the DOLPHOT and KS2 magnitudes over a magnitude range selected to omit bright saturated stars and faint noisy stars (Figure \ref{fig:ZP}). The uncertainty on the average difference is less than 0.1\% for all filters, and so the KS2 zeropoint uncertainty in dominated by the reported HST zeropoint uncertainty of 1\% \citep{Kalirai:2009qf}. The final zeropoints (in magnitudes) are 32.6783, 25.2305, 23.566, 23.088, and 24.5698 for the F814W, F125W, F127M, F153M, and F160W filters, respectively.

The proper motions provide a reliable method to separate likely cluster members from field stars. Following H15, a Gaussian Mixture model was used to describe the kinematic distributions of the cluster and field, from which a cluster membership probability is calculated for each star. The resulting proper motion catalog contains 9922 stars with membership probabilities and is presented in detail in Lu et al., in prep.

\begin{deluxetable*}{llllrrccrrc}
\tablewidth{0pt}
\tabletypesize{\scriptsize}
\tablecaption{Observations}
\tablehead{
\colhead{Date} &
\colhead{Target} &
\colhead{Filter} &
\colhead{Telescope/Inst} &
\colhead{P.A.} &
\colhead{t$_{exp}$\tablenotemark{a}} &
\colhead{N$_{\textrm{img}}$\tablenotemark{b}} &
\colhead{N$_{\textrm{mosaic}}$\tablenotemark{c}} &
\colhead{N$_{\textrm{stars}}$} &
\colhead{Depth\tablenotemark{d}}
\\
\colhead{} &
\colhead{} &
\colhead{} &
\colhead{} &
\colhead{(deg)} &
\colhead{(s)} &
\colhead{} &
\colhead{} &
\colhead{} &
\colhead{(mag)} &
}
\startdata
2005.485 & Wd1 & F814W & \emph{HST} ACS-WFC &46.43 & 802 &  3 & 1          & 10,056 & 24.3  \\
2010.652 & Wd1 & F125W & \emph{HST} WFC3-IR  & -45.87 & 349 &  7 & 2$\times$2 & 10,029 & 22.1  \\
2010.652 & Wd1 & F160W & \emph{HST} WFC3-IR  & -45.87 & 299 &  7 & 2$\times$2 & 10,056 &  20.8 \\
2013.199 & Wd1 & F160W & \emph{HST} WFC3-IR & 134.67 & 299 & 14 & 2$\times$2 & 10,056 & 20.8 \\
2010.351\tablenotemark{e} & Wd1 & K$_s$ & \emph{VISTA} VVV & 49.5 & 4    &    4 &             &  5990   &  15.8 \\
2010.615 & Arches & F127M & \emph{HST} WFC3-IR & -45.33  & 600 & 12 & 1 &   30,530  & 21.8 \\
2010.604 & Arches & F153M &\emph{HST} WFC3-IR  & -45.33  & 350 & 21 & 1 &   30,530 & 20.5  \\
2011.683 & Arches & F153M & \emph{HST} WFC3-IR & -45.33  & 350 & 21 & 1 &   30,530 &  20.5 \\
2012.616 & Arches & F153M & \emph{HST} WFC3-IR & -45.33  & 350 & 21 & 1 &   30,530  & 20.5 \\
\enddata
\tablenotetext{a}{Exposure time for a single image}
\tablenotetext{b}{Number of images at each dither position}
\tablenotetext{c}{Mosaic pattern used in observations}
\tablenotetext{d}{Magnitude at which median error is 0.05 mags}
\tablenotetext{e}{File: ADP.2014-11-25T14:28:20.543.fits}
\label{tab:HST_obs}
\end{deluxetable*}

\subsection{VISTA Observations: Wd 1}
\label{sec:Wd1VISTAobs}
Using the 4m \emph{VISTA} telescope at Cerro Paranal, the \emph{VVV} survey mapped 520 deg$^2$ of the Milky Way bulge and disk in the K$_s$-band \citep{Minniti:2010lp}. This area was observed in 1.64 deg$^2$ tiles, each composed of 6 individual pointings dithered such that each pixel (except for those at the extreme edges of the tile) was covered by at least 4 images. The \emph{VVV} photometric catalog for the tile containing Wd1 was downloaded from Data Release 2 using the ESO Phase 3 Archive Interface\footnote{http://archive.eso.org/wdb/wdb/adp/phase3\_vircam/form}. However, the positions in the catalog revealed that there were only limited detections near Wd1, likely due to significant stellar crowding in the field. In addition, the \emph{VVV} catalog contains aperture photometry, which can be compromised in crowded regions such as Wd1. As a result, we performed point spread function (PSF) photometry on the \emph{VISTA} tile directly to improve the measurements in the Wd1 field.

After downloading the VISTA K$_s$ tile image using the Phase 3 Archive Interface, we trimmed the tile to a manageable region (4.2' x 4.2') that overlaps our \emph{HST} observations. To perform PSF photometry we used \emph{AIROPA}, an expansion of the \emph{Starfinder} code \citep{Diolaiti:2000rc}, in the legacy mode \citep{Witzel:2016lq}. In this mode, \emph{AIROPA} behaves identically to \emph{Starfinder} v1.6. Briefly, an empirical PSF is derived from a subset of user-identified stars and then cross-correlated with the image to detect stars above a user-defined threshold. We carefully selected 27 relatively isolated and non-saturated stars in the tile to define the model PSF and set the minimum correlation coefficient of 0.7. This results in the detection of $\sim$6000 stars in the image.

We calibrate the PSF photometry by matching sources between the PSF and \emph{VVV} catalogs. A two-dimensional first-order polynomial is used to transform the \emph{VVV} positions into the PSF catalog astrometric reference frame, obtaining 1300 stars with positions matched within 1" (the FWHM of the image). Only stars with photometric errors less than 0.05 mags in both catalogs and fainter than the saturation limit are considered. Further, we require each star's \emph{VVV} 1" aperture diameter magnitude (``APERMAG1'') be consistent with its 2" aperture diameter magnitude (``APERMAG3'') to within 0.05 mags, in an effort to eliminate stars with close neighbors affecting the aperture photometry. After these cuts 136 stars remain. The photometric zeropoint for the PSF catalog is calculated from the median difference between the \emph{VVV} magnitude and the instrumental PSF magnitude (Figure \ref{fig:ZP}). The final K$_s$ zeropoint is 24.566 $\pm$ 0.012 mags, where the uncertainty is the median difference error (0.005 mags) combined in quadrature with the \emph{VVV} zeropoint error reported for the tile image (0.011 mags).

It has been shown that the photometric errors reported by \emph{Starfinder} (and thus \emph{AIROPA} in single-PSF mode) are systematically underestimated because they do not capture errors in the PSF model itself \citep{Schodel:2010kx}. This PSF uncertainty is largely constant with magnitude, usually dominating the error budget for bright stars where the photon noise is very low. We derive this additional error term from the same sample of matched stars used to derive the photometric zeropoint. We assume that the standard deviation of the difference between the PSF and \emph{VVV} catalog magnitudes ($\sigma_{\Delta}$) is a combination of the\emph{VVV} catalog error ($\sigma_{VVV}$), PSF error reported by \emph{AIROPA} ($\sigma_{A}$), and the constant PSF error term ($\sigma_{PSF}$):

\begin{equation}
\sigma_{\Delta}^2 = \sigma_{VVV}(m)^2 + \sigma_{A}(m)^2 + \sigma_{PSF}^2
\end{equation}

where $\sigma_{VVV}$ and $\sigma_{A}$ are both functions of the magnitude $m$. We find $\sigma_{PSF}$ = 0.058 mag, which is added in quadrature with the reported \emph{AIROPA} errors to produce the final photometric errors for the PSF catalog.

\subsection{HST Observations: Arches Cluster}
The Arches cluster was observed with \emph{HST} WFC3-IR using the F127M and F153M filters in 2010 (GO-11671; PI: A.M. Ghez), and then repeat observations in F153M were obtained in 2011 and 2012 for astrometric purposes (GO-12318, GO-12667; PI: A.M. Ghez). These observations, along with the photometric and astrometric measurements extracted using KS2, and are presented in H15. The photometric zeropoints are derived in the same manner as the Wd1 \emph{HST} observations and are also shown in Figure \ref{fig:ZP}. We use the same catalog presented in H15, which contains $\sim$26,000 stars.

\section{Methods}
\label{sec:Methods}
We forward-model the Wd1 stars and Arches field RC photometry, simultaneously allowing the extinction law, Wd1 total extinction and distance, and RC star average distance and magnitude spread to vary in order to achieve the best-fit. The extinction law can be divided into two components: the wavelength-dependent extinction law shape and the wavelength-independent normalization factors (see Appendix \ref{app:defs}). This analysis assumes that both components of the NIR extinction law (JHK) are the same along the Wd1 ($\ell$ = -20.451$^{\circ}$, $b$ = -0.404$^{\circ}$) and Arches cluster ($\ell$ = 0.121$^{\circ}$, $b$ = 0.0168$^{\circ}$) lines of sight (LOS).

While the extinction law shape is known to vary across different sightlines in the optical and UV \citep[e.g.][]{Cardelli:1989qf}, the NIR shape has been observed to be relatively constant as a function of total extinction and galactic longitude \citep{Wang:2014xr, Schultheis:2015di, Majaess:2016fk}. A notable exception is the extinction law of \citet{Fitzpatrick:2009ys}, which exhibits significant variations in the NIR for a small sample of targets. However, \citet{Schlafly:2016cr} derive a law for a large set of sightlines that calls for minimal variation in the NIR, supporting the case for a similar NIR shape for the Wd1 and Arches LOS. We confirm this assumption by comparing the shape of Wd1-data only and Wd1+RC extinction law fit in $\mathsection$\ref{sec:Wd1Archesresults}.

The normalization of the extinction law is more challenging to measure since it requires knowledge of the distance to the source object. As a result, how the normalization might change for different LOS is not well studied. However, the extinction along the Wd1 and Arches LOS are dominated by similar material, namely dust from foreground spiral arms in the Galactic Plane. Further, it is not clear why the normalization of the law would be different if the shape is the same. We move forward assuming that the normalization factors are the same for both clusters, though this will require future verification ($\mathsection$\ref{sec:future}).

A Bayesian analysis is used to compare a given extinction law model to the observations. The likelihood function contains a component for the Wd1 MS stars and a component for the RC stars, which are combined in the final analysis. In this section, $\mathsection$\ref{sec:Iso} describes the stellar models used to simulate the observed populations, $\mathsection$\ref{sec:Wd1samp} and $\mathsection$\ref{sec:RCsamp} describe the observed Wd1 and Arches RC samples, and $\mathsection$\ref{sec:fitter} describes the extinction law model, likelihood equation, and subsequent tests of the analysis.

\subsection{Stellar Models and Synthetic Photometry}
\label{sec:Iso}
To model the Wd1 and RC observations, we use stellar models to represent the stellar population, apply extinction to their spectra using a generated extinction law, and then calculate synthetic photometry in the observed filters. To generate the Wd1 stars we must first adopt a cluster age. Spectroscopic studies of the evolved star population suggest a cluster age between 4 -- 6 Myr, based on the properties of the O-type supergiants \citep{Negueruela:2010hc} and the observed ratio of different populations of supergiants to Wolf-Rayet stars \citep{Crowther:2006hb}. Analysis of the pre-main sequence (pre-MS) turn-on feature in the CMD has found cluster ages between 3 -- 5 Myr \citep{Brandner:2008eu, Gennaro:2011nx}, but the age is degenerate with the cluster distance \citep[e.g.][]{Andersen:2017aq}, which is uncertain to $\pm$ 700 pc \citep[$\sim$18\%;][]{Kothes:2007jx}. None of these studies find evidence of an age spread in the cluster, and \citet{Clark:2005sp} point out that the large number of massive stars would be expected to remove excess gas in the cluster within a crossing time. We thus assume the cluster is coeval and adopt an age of 5 Myr. We will perform separate analyses assuming ages of 4 Myr and 6 Myr to assess the impact of the age uncertainty on the extinction law.

A theoretical Wd1 cluster isochrone is generated with the adopted age and assuming solar metallicity using Geneva evolution models with rotation \citep[$\Omega$ = 0.4;][]{Ekstrom:2012qm} for the main sequence/evolved stars and Pisa evolution models \citep{Tognelli:2011fr} for the pre-MS. For each star in the isochrone, the effective temperature, $T_{eff}$, and surface gravity, $\log(g)$, from the stellar evolution model are used to select a model atmosphere. The atmospheres are drawn from a combined library of ATLAS9 \citep{Castelli:2004yq} and PHOENIX \citep[version 16;][]{Husser:2013ts} synthetic spectra. ATLAS9 spectra are used for T$_{eff} >$ 5500 K and PHOENIX spectra are used for T$_{eff} <$ 5000 K, with a linear interpolation of the two model sets between 5000 -- 5500 K. The spectra are then reddened using a custom-built extinction law ($\mathsection$\ref{sec:buildExtinctionLaw}) and total extinction using \emph{Pysynphot} \citep{STScI-Development-Team:2013fd}. The reddened spectra are convolved with the desired filter functions to produce synthetic photometry. The output magnitudes can then be scaled to any cluster distance (a free parameter in our model) for a direct comparison to the observations.

\begin{figure*}
\begin{center}
\includegraphics[scale=0.45]{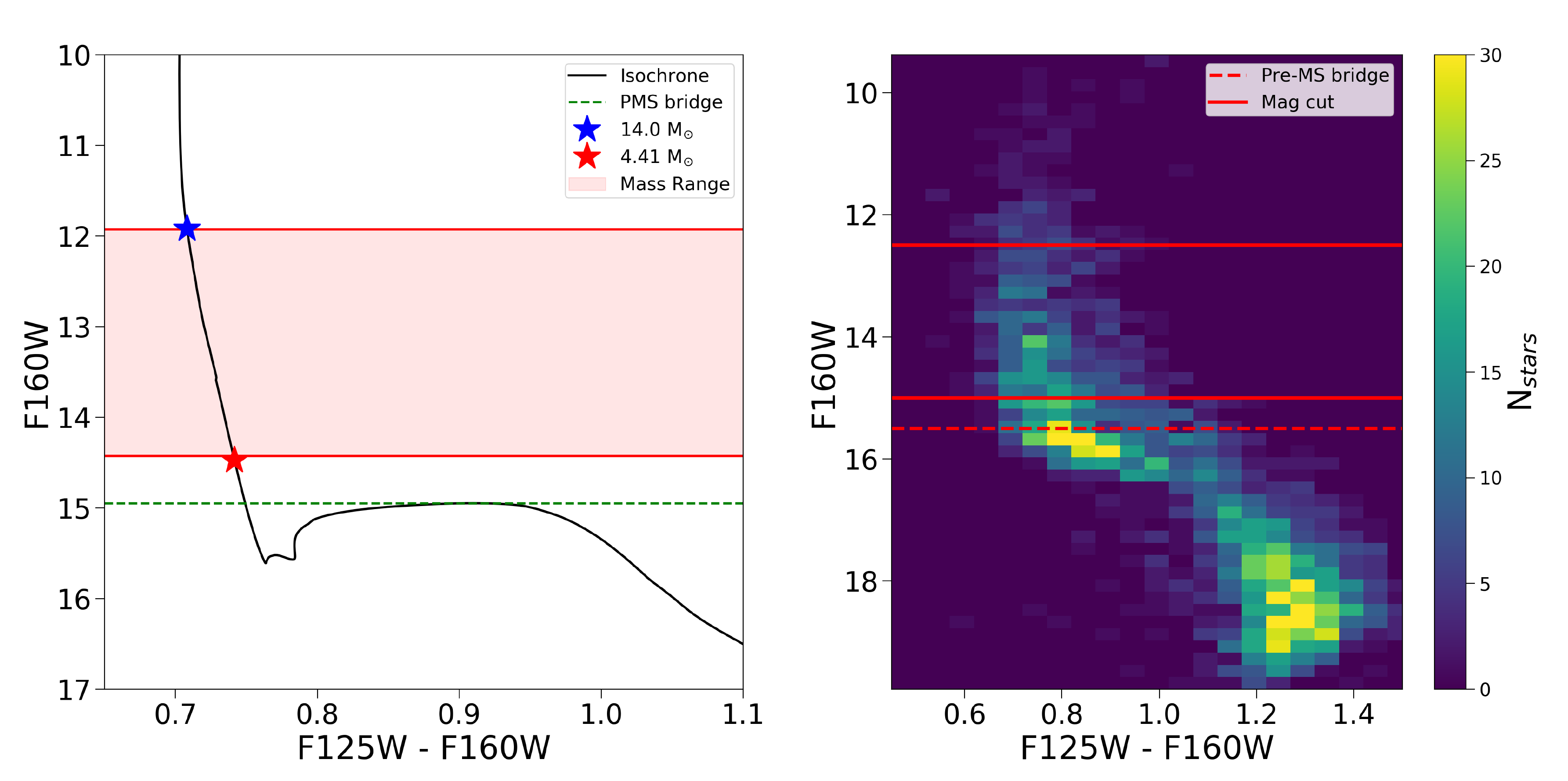}
\end{center}
\caption{The criteria used to select main-sequence cluster members based on a theoretical isochrone. \emph{Left:} The mass range for a 5 Myr isochrone, which is selected based on magnitude relative to the pre-MS bridge (green dashed line). These criteria match those imposed on the observed sample. \emph{Right:} Color-magnitude diagram of proper-motion selected Wd1 members (cluster membership probability P$_{clust}$ $\geq$ 0.6). We identify MS stars in our Wd1 sample as those between 0.5 mags brighter and 3 mags brighter (red lines) than the pre-MS bridge (red dotted line) in F160W.
\label{fig:mass_cut}
}
\end{figure*}

As discussed in $\mathsection$\ref{sec:Wd1samp}, we limit the observed Wd1 sample to only stars that fall on the main sequence (MS). These stars represent an ideal sample to examine the extinction law because their intrinsic colors are well understood and are only slightly dependent on mass in the filters used in this study. We restrict the stellar mass range of the theoretical isochrone to match the selection criterion imposed on the observed sample, selecting stars between 0.5 and 3.0 mags brighter than the pre-MS bridge in F160W (Figure \ref{fig:mass_cut}). The pre-MS bridge, which connects the MS and pre-MS sequences in a color-magnitude diagram (CMD), is only dependent on the cluster age. Thus, no assumptions regarding the cluster distance, extinction law, or total extinction are required in order to appropriately set the isochrone mass range. For a 5 Myr cluster this mass range is  4.41 M$_{\odot}$  -- 14.0 M$_{\odot}$. To test whether the ATLAS9 atmospheres (which assume local thermodynamic equilibrium) are appropriate for the high-mass stars in our sample, we compare the synthetic magnitudes for a 9 M$_{\odot}$, 12 M$_{\odot}$, and 15 M$_{\odot}$ main sequence star with those calculated using non-local thermodynamic equilibrium CMFGEN atmospheres from \citet{Fierro:2015sh}. We find that the synthetic magnitudes agree to within the photometric errors, and so we move forward with the ATLAS9 models in our analysis.

A similar procedure is used to model the Arches field RC stars. We select the RC star model from a 10 Gyr PARSEC stellar isochrone \citep{Bressan:2012ya} at solar metallicity, which represents the average stellar population in the Galactic bulge \citep{Zoccali:2003dw, Clarkson:2008hw}. The chosen model matches the average effective temperature (T$_{eff}$) and surface gravity ($\log(g)$) measured for solar-metallicity RC stars in the \emph{Hipparcos} catalog \citep[T$_{eff}$ = 4700 K, log $g$ = 2.40 cgs;][]{Mishenina:2006ij}. The synthetic photometry is then scaled to the average distance of the RC population, which is a free parameter in our model.

\subsection{Wd1 Sample: Main Sequence Stars}
\label{sec:Wd1samp}
We select a sample of high-probability Wd1 members that fall on the MS, are not saturated, and have small photometric errors. Each star in the sample has photometry in each \emph{HST} filter, and a subset of the stars have \emph{VISTA} K$_s$ photometry as well. The steps used to create the sample are described below.

The sample is created directly from the \emph{HST} proper motion catalog described in $\mathsection$\ref{sec:Wd1HSTobs} and Lu et al., in prep. First, we restrict the catalog to stars with cluster membership probabilities greater than 0.6 and photometric errors less than 0.05 mags in each filter. MS stars are identified as those 0.5 mags brighter than the pre-MS bridge in F160W, corresponding to F160W $\leq$ 15.0 mags (Figure \ref{fig:mass_cut}). This conservative criterion is adopted to minimize contamination from pre-MS stars scattered into the MS by differential reddening. At the bright end, we adopt a magnitude cut of F160W $\geq$12.5 mag in order to eliminate saturated sources in the \emph{VISTA} K$_s$ observations. After these cuts, 537 of the original 9922 stars remain in the sample.

To eliminate photometric outliers (such as field stars or binary systems with unusual colors), we apply an iterative 3-sigma cut in a two-color diagram (2CD) across the HST filters: F814W - F125W vs. F814W - F160W. We fit a line to the sample (via orthogonal regression) and calculate the root-mean-squared (RMS) residual relative to the fit in 0.25 magnitude bins. Stars with residuals larger than 3 times the RMS value in their magnitude bin are removed. This process is repeated until no further stars are eliminated. A total of 53 stars are rejected by this criterion.

\begin{figure}
\begin{center}
\includegraphics[scale=0.35]{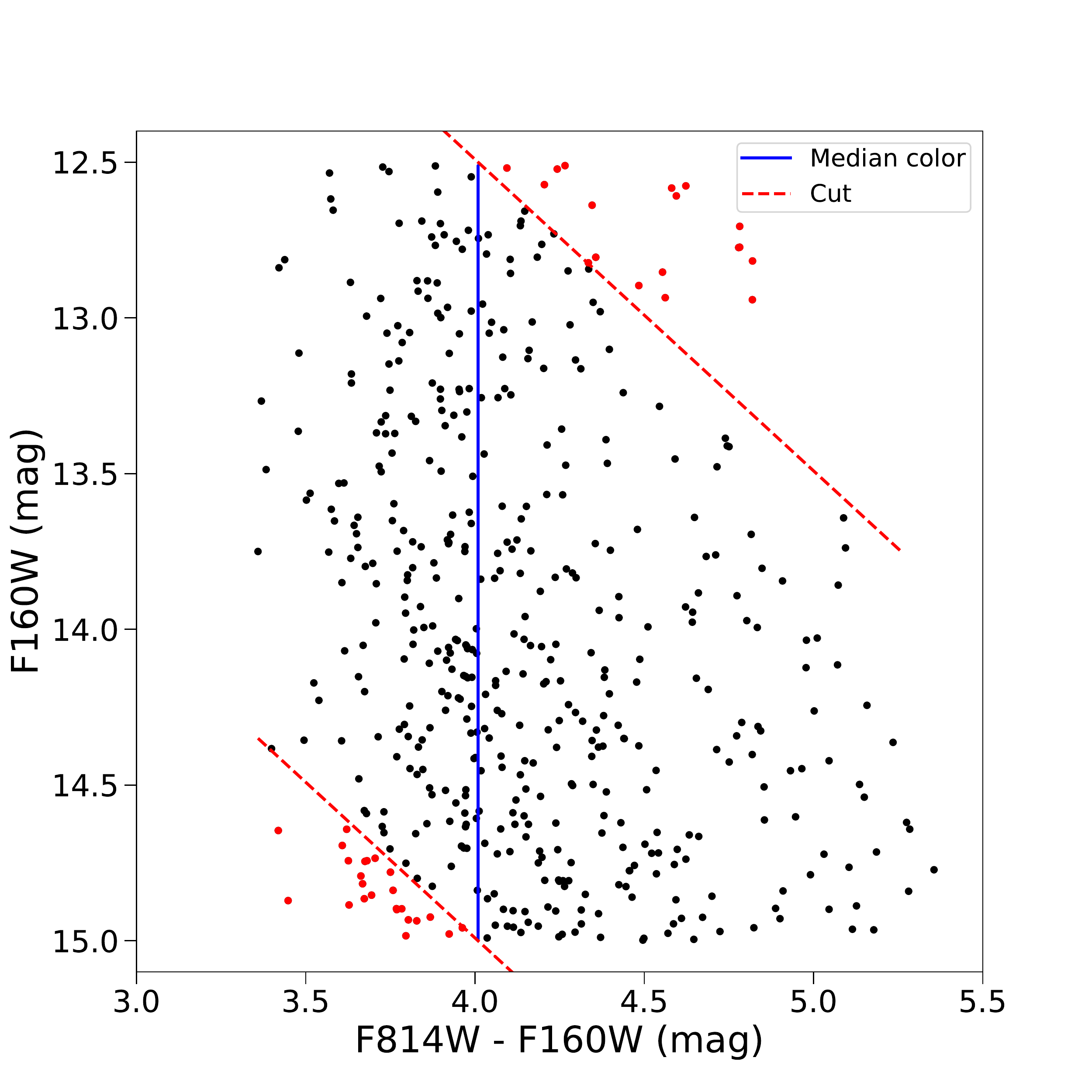}
\end{center}
\caption{Cut applied to the Wd1 sample to eliminate intrinsically brighter/fainter stars that have scattered into the target magnitude range due to differential extinction. Bright (i.e. high mass) interlopers can be scattered into the sample redward of the median cluster sequence while faint (i.e. low mass) interlopers can be scattered blueward. The median cluster sequence is shown by the blue line, the applied cuts by the red dotted lines, and the stars removed by this cut by the red points.
\label{fig:outlier_cut}
}
\end{figure}

A final cut is required to eliminate cluster stars that would otherwise be outside the adopted F160W magnitude range but have been scattered into the sample by differential extinction. For example, a high-mass star intrinsically brighter than the F160W magnitude limit can scatter into the sample if it is located in a region of higher extinction. This star would appear redward of the average main-sequence population. Similarly, a low-mass star could scatter into the sample if it is in a region of lower extinction, placing it blueward of the average main sequence population. To eliminate these stars we make a cut in F160W vs. F814W - F160W CMD space, where the median color is taken to represent the MS. Two lines with a slope of 1 that intersect the bright and faint ends of the MS are used to identify and remove potentially scattered stars (Figure \ref{fig:outlier_cut}). This slope is a conservative estimate of the steepest possible reddening vector in the CMD, corresponding to A$_{F814W}$ / A$_{F160W}$ $\sim$ 2. A steeper reddening vector would require A$_{F814W}$ / A$_{F160W}$ $<$ 2, which is much lower than any law reported in the literature. An additional 42 stars are removed by this criterion, resulting in a final sample size of 453 stars.

To construct the \emph{HST}-\emph{VISTA} subsample, we iteratively match the \emph{HST} proper motion catalog with the \emph{VISTA} catalog using a 0.25Ó ($\sim$2 \emph{HST} HST pixels) matching radius. An additional ``isolation cut'' is imposed to reduce the impact of stellar crowding on the seeing-limited \emph{VISTA} photometry: stars lying within 4.5'' of another star that has a brightness within 3 magnitudes of the star itself (as identified from the \emph{HST} catalog) are rejected. All of the same cuts are applied for a final \emph{HST}-\emph{VISTA} subsample of 106 stars.

\begin{figure*}
\begin{center}
\includegraphics[scale=0.35]{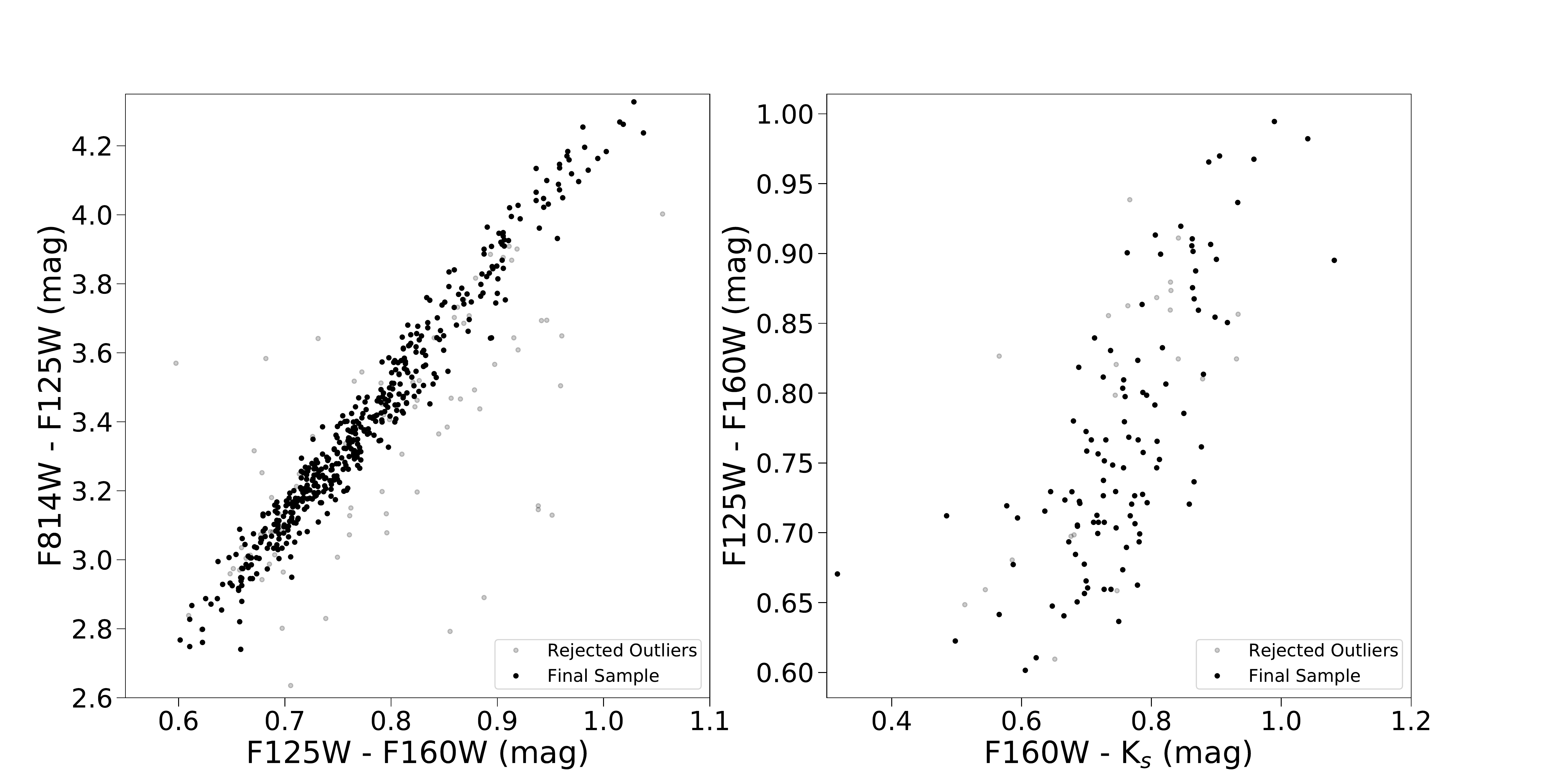}
\end{center}
\caption{Two-color diagrams of final Wd1 sample, with the \emph{HST}-only 2CD on the left and \emph{HST}-\emph{VISTA} 2CD to the right. Stars rejected as photometric outliers or scattered stars are shown in gray, while the final sample is shown in black. The high photometric precision of \emph{HST} relative to \emph{VISTA} is evident by the scatter introduced by the \emph{VISTA} photometry in the \emph{HST}-\emph{VISTA} 2CD.
\label{fig:Wd1_tcd}
}
\end{figure*}

A summary of the adopted cuts and sample size is provided in Table 2. These stars form well-defined reddening vectors in the \emph{HST}-only and \emph{HST}-\emph{VISTA} 2CDs, which are used in the extinction law analysis (Figure \ref{fig:Wd1_tcd}). The typical photometric uncertainty is $\sim$0.02 mag for the \emph{HST} filters and $\sim$0.07 mag for \emph{VISTA} K$_s$. While the magnitude cuts introduce a possible Malmquist bias, the effect is small due to the small photometric errors. Further, a bias would only affect our results if the extinction law for the faint stars were different than that of the bright stars, which is highly unlikely since they are part of the same cluster.

\begin{deluxetable*}{l l c c}
\label{tab:cuts}
\tablecaption{Wd1 Sample Selection}
\tabletypesize{\scriptsize}
\tablehead{
\colhead{Selection Description} & \colhead{Selection Criterion} & \colhead{N$_{stars}$ cut from HST Sample} & \colhead{N$_{stars}$ cut from HST-VISTA Sample}
}
\startdata
{\bf Original Sample Size} &    & {\bf 9922}  & {\bf 1071}  \\
Membership & P$_{clust} \geq 0.6$ & 7007 & 426 \\
Phot Error & $\leq$ 0.05 mag & 826 &  15 \\
Phot Range &  12.5 $\leq$ F160W $\leq$ 15.0 & 1552 & 272 \\
Isolation & 4.5", $\geq$ star mag + 3 & ---  & 229 \\
Sigma-clipping & 3$\sigma$, iterative &  42 & 12 \\
High/low-mass scatter & see $\mathsection$\ref{sec:Wd1samp} & 42  & 11 \\
{\bf Final Sample Size} &    & {\bf 453}  & {\bf 106}  \\
\enddata
\end{deluxetable*}

\subsection{Arches sample: Red Clump Stars}
\label{sec:RCsamp}
RC stars are low-mass giants that are in the core helium-burning stage of their evolution. Exhibiting a narrow range of T$_{eff}$ and $\log(g)$ values, these stars form a well defined clump in the CMD that spreads along the reddening vector in the presence of differential extinction \citep[][for review]{Girardi:2016ho}. While no significant RC star population is found in the Wd1 observations, H15 found a large RC field population in NIR \emph{HST} observations of the Arches cluster. These stars are associated with the Galactic Bulge and have a distance distribution that peaks close to the GC along this sight-line. Since the GC distance is known to 2\% \citep{Boehle:2016rt}, the normalization of the extinction law to these stars can be constrained to a much higher precision than is possible with Wd1.

\begin{figure*}
\begin{center}
\includegraphics[scale=0.35]{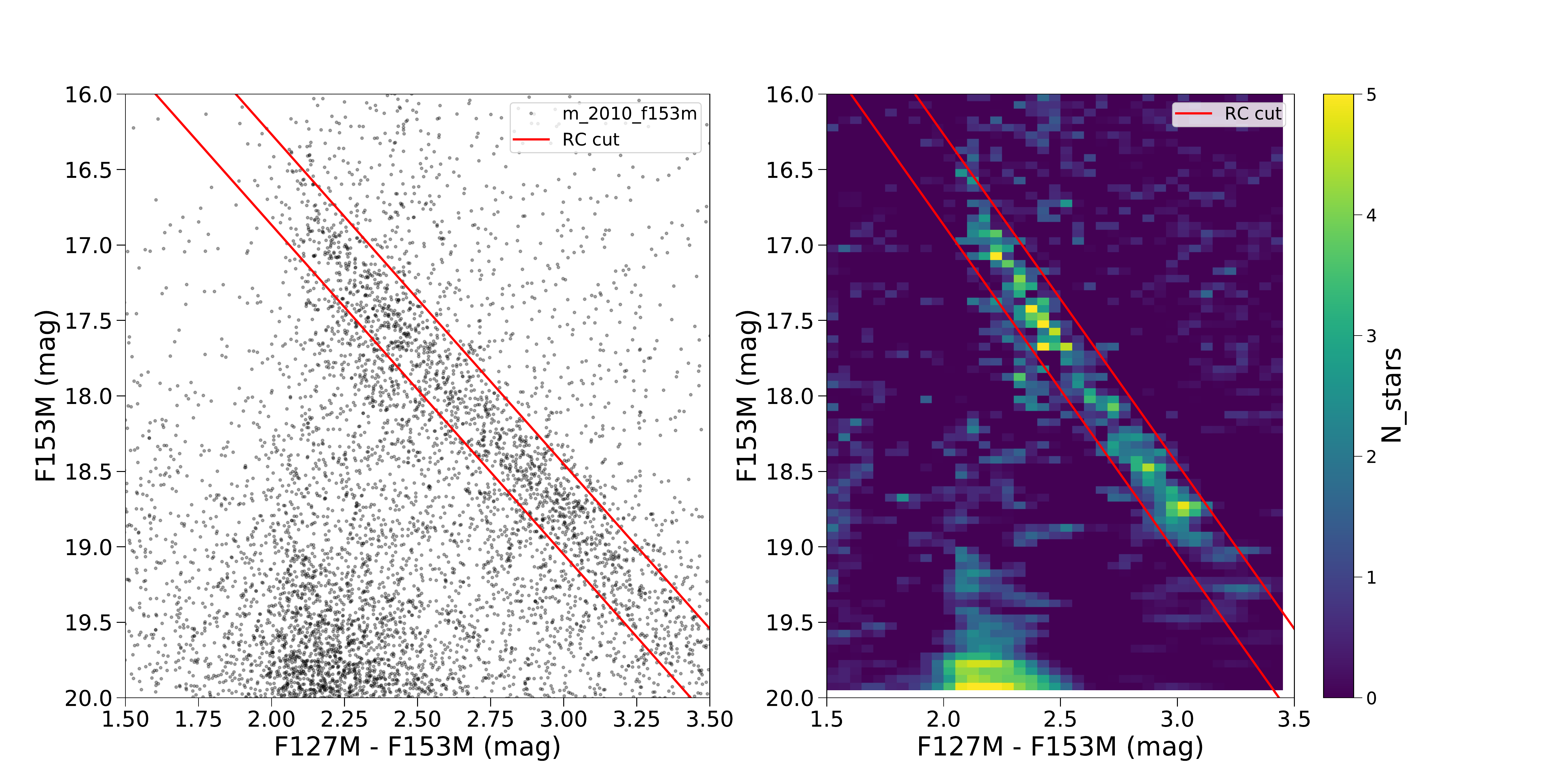}
\end{center}
\caption{The selection criterion used to identify RC stars in the Arches cluster field. Following \citet{De-Marchi:2016qw}, we convert the observed CMD (left) into a Hess diagram and perform unsharp-masking to identify the high-density ridge corresponding to the RC population. The resulting unsharp-masked density histogram is shown to the right. RC stars are identified as those falling within +/- 0.3 mags of a linear fit to the RC density histogram. This selection criterion is shown by the red lines.
}
\label{fig:RCcut}
\end{figure*}

RC stars are visible in the Arches field CMD as a high-density ``bar'' (Figure \ref{fig:RCcut}). To identify the RC stars we use the unsharp-masking technique described by \citet{De-Marchi:2016qw}. This method increases the contrast of high-frequency features while reducing the contrast of low-frequency ones. First, we convert the F153M vs. F127M - F153M CMD into a Hess diagram (i.e., a 2D histogram of stellar density), treating the position of each star as a Gaussian probability distribution with a width corresponding to the photometric error. A bin size of 0.05 mags in both both color and mag space is used. Next, we create the unsharp mask by convolving the Hess diagram with a 2D Gaussian kernel having a width of 0.2 mags. The mask is then subtracted from the original Hess diagram to create the unsharp-masked diagram. We calculate a linear fit to the high-density ``ridge'' created by the RC population and identify RC stars as those that fall within F153M +/- 0.3 mags of the best-fit line (Figure \ref{fig:RCcut}). A total of 1119 RC stars are identified in this manner.

Similar to the Wd1 sample, we adopt a series of cuts to ensure the quality of the RC sample for the extinction law analysis. We require each star to have a photometric error less than 0.05 mag in both filters and perform a 3$\sigma$ iterative outlier rejection in the CMD. The final RC sample contains 819 stars with typical photometric uncertainties of 0.015 mags.

\begin{figure}
\begin{center}
\includegraphics[scale=0.35]{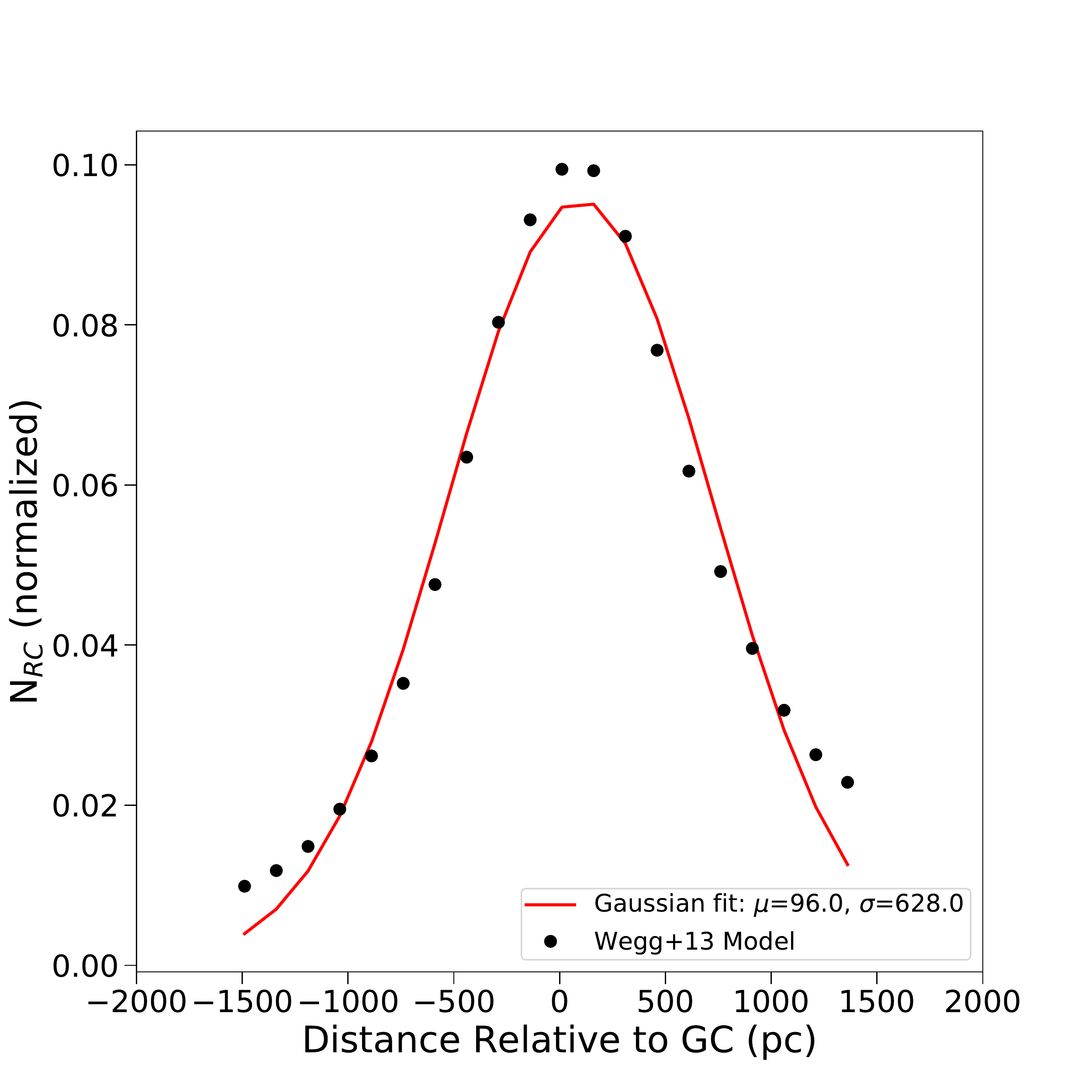}
\end{center}
\caption{The expected distribution of RC stars as a function of distance in the Arches field based on RC density curves of \citet{Wegg:2013pd}. The distribution (black points) is approximated by a Gaussian (red line) centered 96 pc beyond the GC itself with a width of 630 pc.
}
\label{fig:RCdens}
\end{figure}

The expected distance distribution of the sample is calculated using the RC density profiles of \citet{Wegg:2013pd}, which are measured for the major, intermediate, and minor axes of the Galactic bar. These profiles are rotated by 28$^{\circ}$ \citep[the measured angle of the Galactic bar relative to the GC;][]{Wegg:2013pd} to determine the density profile of the Arches LOS. The expected number of RC stars is then calculated as a function of distance by multiplying the solid angular area of the observations by the LOS density profile. The resulting distribution is nearly Gaussian with a peak at 96 pc beyond the GC and a width of 630 pc (Figure \ref{fig:RCdens}). The peak in the observed counts is not centered at the GC due to the increase in projected field area with LOS distance.

The RC star distance distribution is robust against uncertainties in the Galactic bar rotation angle ($\alpha$) and the RC density profiles. Varying $\alpha$ between 25$^{\circ}$ -- 33$^{\circ}$ (the possible range reported by \citeauthor{Wegg:2013pd} 2013 and \citeauthor{2015MNRAS.450.4050W} 2015) only shifts the peak of the distribution by 12 pc. Redrawing the RC density profiles of \citet{Wegg:2013pd} while applying the reported uncertainty of 10\% on each measurement results in a typical shift in the distribution of just $\pm$20 pc. Both sources of error are well within the uncertainty in the GC distance itself (140 pc) and are not considered in the analysis.

\begin{figure*}
\begin{center}
\includegraphics[scale=0.45]{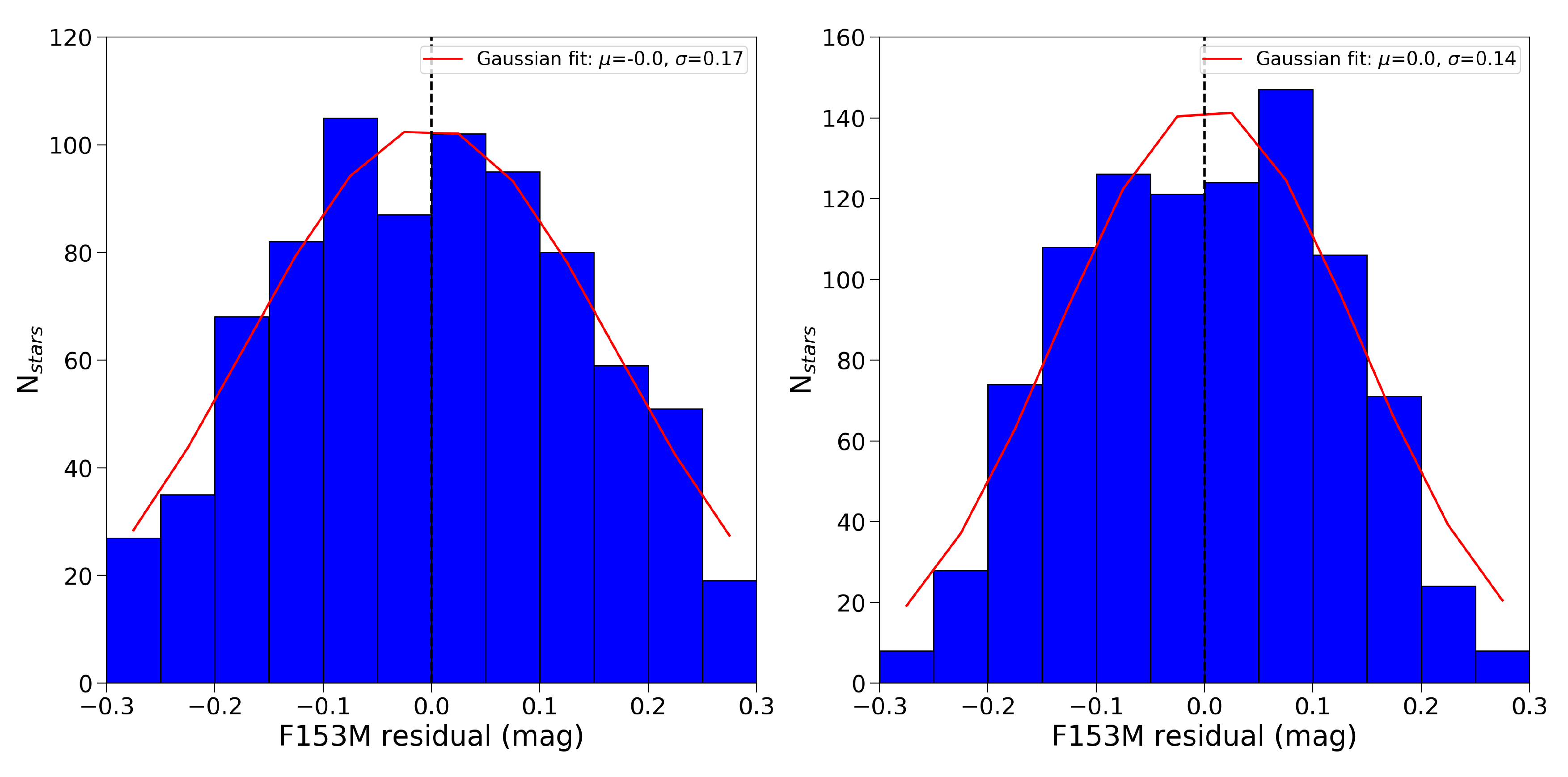}
\end{center}
\caption{The distribution of F153M residuals in CMD space for the observations (left) and synthetic RC stars with $\sigma_d$ = 630 pc (right). The observed distribution is wider, likely caused by intrinsic magnitude variations within the observed RC population due to differences in stellar age and/or metallicity.}
\label{fig:RCresid}
\end{figure*}

In the CMD, we expect the distribution of F153M residuals relative to the RC ridge to be driven by the variation in stellar distance across the population. To test this, we compare the observed residuals to those calculated for a synthetic RC population created with the predicted distance distribution in Figure \ref{fig:RCdens}. Details on how the synthetic population is created is provided in Appendix \ref{app:art}. The observed F153M residual distribution is wider than the residual distribution for the synthetic data by $\sim$0.03 mag (Figure \ref{fig:RCresid}). This indicates that either the RC distance distribution is slightly wider than expected or that there is an additional source of magnitude dispersion among the RC stars. Given that the synthetic RC population assumes a single age and metallicity, it is likely that variations in stellar age or metallicity within the bulge are also impacting the observed F153M residual distribution \citep[e.g.][]{2002MNRAS.337..332S, Chen:2017gf}.

\subsection{Extinction Law Fitter}
\label{sec:fitter}

\subsubsection{Extinction Law Model and Priors}
\label{sec:buildExtinctionLaw}
To construct the extinction law, we define A$_{\lambda}$~/~A$_{Ks}$ at 5 specific wavelengths and use a cubic B-spline to interpolate the law across all wavelengths. The extinction law is thus a continuous function that can be adjusted by changing the A$_{\lambda}$ / A$_{Ks}$ values at the wavelength points, which are free parameters in our model. There are 10 additional free parameters in the model: 4 for the global Wd1 and RC population parameters (Wd1 total extinction A$_{Ks}$; Wd1 distance d$_{wd1}$; average RC distance d$_{rc}$; Gaussian width of F153M residuals around the reddening vector $\sigma_{rc}$), and 6 for systematic offsets in the photometric zeropoints of the filters ($\Delta$ZP$_{\lambda}$). The parameters and adopted priors are discussed below and summarized in Table \ref{tab:ModelParams}.

In the extinction law, a wavelength point is assigned to the pivot wavelength \citep[][]{Tokunaga:2005if} of each Wd1 filter (F814W, F125W, and F160W). The A$_{\lambda}$~/~A$_{Ks}$ values at these points are given uniform priors. Two additional wavelength points are added for the SST/IRAC [3.6] and PanSTARRS $y$ filters, though no observations at these wavelengths are used. Our analysis revealed that the $y$-band point is required to capture the needed curvature in the extinction law between the F125W and F814W filters ($\mathsection$\ref{sec:sensitivity}). We adopt a uniform prior for A$_{\lambda}$~/~A$_{Ks}$ at this point, as well. The [3.6] point is included to enforce reasonable behavior through the red-edge of the extinction law. We adopt a gaussian prior using the value of \citet{Nishiyama:2009fc} with a conservative uncertainty of 10\% (A$_{3.5}$ / A$_{Ks}$ = 0.5 $\pm$ 0.05). The cubic B-spline interpolation is calculated using the \emph{scipy.interpolate.splrep} function in python between 0.8 $\mu$m -- 2.2 $\mu$m. The exact function call is provided in the stand-alone python code referenced in $\mathsection$\ref{sec:Wd1Archesresults}. This is converted into a \emph{pysynphot} custom reddening law for the synthetic photometry. Since each A$_{\lambda}$~/~A$_{Ks}$ point is allowed to vary independently, no assumption regarding the functional form of the extinction law is made.

For the global Wd1 population parameters, we adopt a uniform prior for A$_{Ks}$ and a Gaussian prior of 3900~$\pm$~700 pc for d$_{wd1}$. The distance constraint is derived from the kinematics of HI gas associated with the cluster \citep{Kothes:2007jx}. Though additional distance measurements exist from an eclipsing binary analysis \citep{Koumpia:2012ij}, spectrally-typed evolved stars \citep[e.g.][]{Negueruela:2010hc}, and CMD fitting of pre-MS stars \citep[e.g.][]{Andersen:2017aq}, these analyses must correct for extinction and thus depend on the extinction law.

For the RC stars, we adopt a Gaussian prior of 7960 $\pm$ 140 pc for d$_{rc}$. This is 100 pc beyond the GC distance measurement of \citet[][]{Boehle:2016rt}, matching the predicted average population distance in $\mathsection$\ref{sec:RCsamp}. For $\sigma_{rc}$, we adopt a prior of 0.17 $\pm$ 0.01 mags, corresponding to the measured width found in Figure \ref{fig:RCresid}.

The zeropoint offsets are included in the model since errors in the zeropoints would propagate through the analysis as a systematic error rather than as a random one. The offsets are assigned Gaussian priors centered at zero with a width corresponding to the uncertainty in the zeropoint derivation in $\mathsection$\ref{sec:Wd1HSTobs}.

\begin{deluxetable*}{l l c c c}
\tablewidth{0pt}
\tabletypesize{\footnotesize}
\tablecaption{Model Parameters and Priors}
\tablehead{
\colhead{Parameter\tablenotemark{a}} & \colhead{$\lambda$\tablenotemark{b} ($\mu$m)} & \colhead{Prior\tablenotemark{c}} & \colhead{Units} & \colhead{Prior Reference}
}
\startdata
A$_{F814W}$ / A$_{K_s}$ &  0.806 & U(4, 14) & ---  & ---  \\
A$_{y}$ / A$_{K_s}$ & 0.962 & U(4, 14) & ---  & ---  \\
A$_{F125W}$ / A$_{K_s}$ &  1.25 & U(1, 6) & --- & ---  \\
A$_{F160W}$ / A$_{K_s}$ &  1.53 & U(1, 6) & --- & ---  \\
A$_{[3.6]}$ / A$_{K_s}$ & 3.545 & G(0.5, 0.05) & --- & \citet{Nishiyama:2009fc}  \\
A$_{K_s}$ & 2.14 & U(0.3, 1.3) & mag  & --- \\
d$_{wd1}$ &  --- & G(3900, 700) & pc &  \citet{Kothes:2007jx} \\
d$_{rc}$ &  --- & G(7960, 140) & pc  &  \citet{Boehle:2016rt},  $\mathsection$\ref{sec:RCsamp} \\
$\sigma_{rc}$ &  --- & G(0.17, 0.01) & mag & $\mathsection$\ref{sec:RCsamp} \\
$\Delta$ZP$_{Ks}$ & --- & G(0, 0.012) & mag &   $\mathsection$\ref{sec:Wd1VISTAobs} \\
$\Delta$ZP$_{F160W}$ & --- & G(0, 0.01) & mag  &  $\mathsection$\ref{sec:Wd1HSTobs} \\
$\Delta$ZP$_{F153M}$ & --- & G(0, 0.01) & mag &  $\mathsection$\ref{sec:Wd1HSTobs} \\
$\Delta$ZP$_{F127M}$ & --- & G(0, 0.01) &  mag&  $\mathsection$\ref{sec:Wd1HSTobs} \\
$\Delta$ZP$_{F125W}$ & --- & G(0, 0.01) & mag &  $\mathsection$\ref{sec:Wd1HSTobs} \\
$\Delta$ZP$_{F814W}$ & --- & G(0, 0.01) & mag & $\mathsection$\ref{sec:Wd1HSTobs}  \\
\enddata
\label{tab:ModelParams}
\tablenotetext{a}{A$_{\lambda}$~/~A$_{Ks}$: extinction law in filter; A$_{Ks}$: total extinction of Wd1; d$_{wd1}$: distance to Wd1; d$_{rc}$: average RC star distance; $\sigma_{rc}$: gaussian width of the RC F153M residuals around the reddening vector; $\Delta$ZP$_{\lambda}$: zeropoint offset in filter}
\tablenotetext{b}{HST + PanSTARRS filters: Pivot wavelengths of filter; IRAC [3.6] filter: isophotal wavelength from \citet{Nishiyama:2009fc}}
\tablenotetext{c}{Uniform distributions: U(min, max), where min and max are bounds of the distribution; Gaussian distributions: G($\mu$, $\sigma$), where $\mu$ is the mean and $\sigma$ is the standard deviation}
\end{deluxetable*}

\subsubsection{Wd1 Star Likelihood}
For the Wd1 component of the likelihood, the observed sample is compared to the theoretical isochrone produced by the model in terms of the F160W magnitude and F160W - F814W, F160W - F125W, and F160W - K$_s$ colors. The full sample is used for the \emph{HST} colors and the \emph{HST}-\emph{VISTA} subsample is used for F160W - K$_s$. Since a smaller number of stars have \emph{VISTA} photometry, the \emph{HST} colors have larger weight in the extinction law fit. However, this is desirable as the K$_{s}$ observations have larger scatter than the \emph{HST} observations.

Initially, the observed photometric errors are smaller than the magnitude sampling of the isochrone. We address this by performing a cubic spline interpolation of the isochrone magnitudes as a function of stellar luminosity and resample in steps of 2.5x10$^{-3}$ in log luminosity. This results in a magnitude spacing of 3.7x10$^{-3}$ mags on the isochrone, which is more than 2 times smaller than the typical magnitude error (0.01 mag).  However, there are a handful of stars with mag errors below this threshold. Since finer isochrone sampling significantly increases the computation time of the analysis, we instead add an error floor of 3.7x10$^{-3}$ mags in quadrature to the observations to avoid this problem.

A single isochrone is insufficient to reproduce the data due to the differential extinction (dA$_{K_s}$) in the field. Instead, we generate a grid of isochrones with a range of extinction values $\pm$ 0.6 mags from the input A$_{K_s}$ in steps of 5x10$^{-4}$ mags. This ensures that the color sampling between isochrones is at least 2 times smaller than the photometric uncertainty for each color in the model.

For each observed Wd1 MS star, we identify the nearest-neighbor synthetic star in the multi-dimensional magnitude-color space described above. We define the set of observed magnitudes and colors as $\bm{m}$ and their associated errors $\bm{\sigma}$. The likelihood is:

\begin{equation}
\begin{split}
\mathcal{L}_{Wd1}(\bm{m}, \bm{\sigma}) = \prod_{j=1}^{4} \mathcal{L}_j (\bm{m_{j}}, \bm{\sigma_{j}}) \\
\end{split}
\end{equation}

where $\bm{m_{j}}$ and $\bm{\sigma_{j}}$ are the measurements and errors in the $j$th dimension, with the dimensions corresponding to F160W, F160W~-~F814W, F160W~-~F125W, and F160W~-~K$_{s}$. For each observed star, the nearest neighbor synthetic star across all dimensions is found. The likelihood for each dimension is then calculated by comparing the observed sample to their corresponding nearest neighbors:

\begin{equation}
\mathcal{L}_j(\bm{m_{j}}, \bm{\sigma_{j}}) = \prod_{i=0}^{N_{s,j}}\frac{1}{\sigma_{j,i}\sqrt{2\pi}}e^{\frac{(m_{j,i} - NN_{mod}(m_{j,i}))^2}{2\sigma_{j,i}^2}}
\end{equation}

where $N_{s,j}$ is the number of stars in the $j$th dimension, $m_{j,i}$ and $\sigma_{j,i}$ are the mag/color and corresponding error of the i$th$ star in the $j$th dimension, and $NN_{mod}(m_{j,i})$ is the mag/color of the $i$th star's nearest neighbor synthetic star in the model. The full sample is used for the F160W, F160W~-~F814W, and F160W~-~F125W dimensions, while only the \emph{HST}-\emph{VISTA} subsample is used for the F160W~-~K$_{s}$ dimension.

\subsubsection{RC Star Likelihood}
We calculate the RC star likelihood component in a similar manner, comparing the observed sample to the reddening vector from the extinction law model in color-magnitude space (F153M and F127M - F153M). With $\bm{m_r}$ and $\bm{c_r}$ representing the set of F153M magnitudes and F127M - F153M colors with errors $\bm{\sigma_{m_r}}$ and $\bm{\sigma_{c_r}}$:

\begin{equation}
\mathcal{L}_{RC} =  \mathcal{L}_{m_r}(\bm{m_{r}}, \bm{\sigma_{m_r}}, \sigma_{rc}) * \mathcal{L}_{c_r}(\bm{c_{r}}, \bm{\sigma_{c_r}})
\end{equation}

where $\sigma_{RC}$ is the free parameter in the model corresponding to the Gaussian width of the F153M residuals around the reddening vector. Each component of the likelihood is defined as:

\begin{equation}
\mathcal{L}_{m_r}(\bm{m_{r}}, \bm{\sigma_{m_r}}, \sigma_{rc}) = \prod_{i=0}^{N_r}\frac{1}{\sqrt{\sigma_{m_r,i}^2 + \sigma_{rc}^2}\sqrt{2\pi}}e^{\frac{(m_{m_r,i} - NN_{mod}(c_r,i))^2}{2(\sigma_{m_r,i}^2 + \sigma_{rc}^2)}}
\end{equation}

\begin{equation}
\mathcal{L}_{c_r}(\bm{c_{r}}, \bm{\sigma_{c_r}}) = \prod_{i=0}^{N_r}\frac{1}{\sigma_{c_r,i}^2\sqrt{2\pi}}e^{\frac{(c_{r,i} - NN_{mod}(c_r))^2}{2\sigma_{c_r,i}^2}}
\end{equation}

where $NN_{mod}(c_r, i)$ is the nearest neighbor model star in color space for the $ith$ star and $N_r$ is the total number of RC stars in the sample. Note that the nearest neighbor is only calculated for color space, due to the extra dispersion in the magnitudes as discussed in $\mathsection$\ref{sec:RCsamp}. The extra dispersion is captured by the $\sigma_{rc}$ parameter in the F153M dimension of the likelihood, which manifests as an extra error term in addition to the photometric errors. Since the dispersion is primarily driven by individual RC distance variation, it is not included in the color likelihood term.

\subsubsection{Final Likelihood}
The final likelihood combines the Wd1 and RC likelihood components. For each to have an equal weight we must account for the fact that the RC sample is significantly larger than the Wd1 sample, which causes $\mathcal{L}_{rc} > \mathcal{L}_{wd1}$ regardless of the quality of the fit. After converting to log-likelihood, we scale the RC likelihood component to the same number of stars as the Wd1 sample (N$_{wd1}$):

\begin{equation}
\log{\mathcal{L}_{tot}} = \log{\mathcal{L}_{wd1}} + \log{\mathcal{L}_{r}} * \frac{N_{wd1}}{N_{rc}}
\end{equation}

With the likelihood function defined, we determine the best-fit extinction law model using Bayes' Theorem:

\begin{equation}
\label{eq:bayes}
P( \bm{M} | \Theta) = \frac{\mathcal{L}_{tot} * P(\bm{M})}{P(\Theta)}
\end{equation}

\noindent where $\bm{M}$ is the model with the set of free parameters [$\bm{A_{\lambda}/A_{Ks}}, \bm{\Delta ZP}, A_{Ks},  d_{wd1}, d_{rc}, \sigma_{rc}$](with $\bm{A_{\lambda}/A_{Ks}}, \bm{\Delta ZP}$ representing the set values at 5 and 6 different wavelengths, respectively) and $\Theta$~=~[$\bm{m}$, $\bm{\sigma}$, $\bm{m_{r}}$, $\bm{\sigma_r}$, $\bm{c_r}$, $\bm{\sigma_{c_r}}$] is the set of observations. $P(\bm{M} | \bm{\Theta})$ is the posterior probability of the model given the data, and P($\bm{M}$) is the prior probability on $\bm{M}$.

The posterior probability distributions are calculated using \emph{Multinest}, a nested sampling algorithm shown to be more efficient than Markov Chain Monte Carlo algorithms in complex parameter spaces with multiple modes or pronounced degeneracies \citep{Feroz:2008yu, Feroz:2009lq}. This iterative technique calculates the posterior probability at a fixed number of points in the parameter space and identifies possible peaks, restricting subsequent sampling to the regions around these peaks until the change in evidence drops below a user-defined tolerance level. An evidence tolerance of 0.5 and sampling efficiency of 0.8 are adopted for 400 active points. To run the sampler we use \emph{Pymultinest}, a convenient wrapper module in python \citep{Buchner:2014wa}.

\subsubsection{Testing the Analysis}
To test the performance of the extinction law fitter, we simulate a set of Wd1 MS and Arches RC star observations with known extinction properties and run it through the analysis. A detailed description of the process used to generate the simulated data and the subsequent results are provided in Appendix \ref{app:art}. In summary, a large degeneracy between d$_{wd1}$, A$_{Ks}$, and the extinction law is obtained when the analysis is limited to the Wd1 MS stars, due to the uncertainty in the extinction law normalization. However, when the Wd1 and RC samples are combined, the fitter successfully recovers all of the model parameters to within 1$\sigma$. The fitter is also able to extract the correct extinction law when just the RC star sample is used. These tests validate our methodology.

\section{Results}
\label{sec:results}
We present three fits of the extinction law: one using only the Wd1 MS sample ($\mathsection$\ref{sec:Wd1results}), one using only the Arches RC sample ($\mathsection$\ref{sec:Archesresults}), and one using the combined Wd1 and RC samples ($\mathsection$\ref{sec:Wd1Archesresults}). The Wd1-only fit constrains the shape of the extinction law, leveraging the large wavelength coverage of the observations. The RC-only fit constrains A$_{F125W}$ / A$_{F160W}$, which defines the normalization. The Wd1~+~RC fit combines the strengths of both data sets to produce the final extinction law.

To quantify the extinction law shape we introduce the parameter $\mathcal{S}_{1/\lambda}$, which is the ratio of the derivative of the B-spline interpolated extinction law with respect to 1/$\lambda$ to the derivative of the law with respect to 1/2.14 $\mu$m:

\begin{equation}
\label{eq:shape}
S_{1/\lambda} = \frac{\frac{\partial(A_{\lambda} / A_{Ks})}{\partial(1/\lambda)}}{\frac{\partial(A_{\lambda} / A_{Ks})}{\partial(1/\lambda)} |_ {\lambda = 2.14 \mu m}}
\end{equation}

The advantage of this quantity over a color excess ratio is that it is continuous as a function of wavelength \citep[as the derivative of a cubic B-spline must be continuous if the knots are distinct;][]{de-Boor:1978mq}, making it easier to compare extinction laws measured in different filters. Further discussion of $\mathcal{S}_{1/\lambda}$ and its relationship to the color excess ratio can be found in Appendix \ref{app:defs}. A summary of the results are presented in Table \ref{tab:results} and selected posterior distributions are shown in Appendix \ref{app:posteriors}. Unless otherwise specified, we report two errors for the parameters in the text below, the first being the statistical error and the second being the systematic error.

\begin{deluxetable*}{l | c c c c | c c c c c c}
\tablewidth{0pt}
\tablecaption{Extinction Law Results}
\tabletypesize{\scriptsize}
\tablehead{
& \multicolumn{3}{c}{This study\tablenotemark{a}} & \multicolumn{6}{c}{Literature\tablenotemark{b}} \\
\colhead{Parameter} & \colhead{Wd1 only} & \colhead{RC only} & \colhead{Wd1 + RC} & \colhead{$\sigma_{sys}$\tablenotemark{c}}  & \colhead{C89} & \colhead{D16} & \colhead{N09} & \colhead{F09} & \colhead{S16$_1$} & \colhead{S16$_2$}
}
\startdata
\cutinhead{Free Parameters}
A$_{Ks}$ (mag) & 0.78 $\pm$ 0.16 & --- & 0.611 $\pm$ 0.024 & 0.02 & --- & 0.74 $\pm$ 0.08 & 0.87 $\pm$ 0.01\tablenotemark{d} & --- & --- & ---\\
d$_{wd1}$ (pc) & 4428 $\pm$ 309 & --- & 4780 $\pm$ 76 & 48 & --- & --- & --- & --- & ---& ---\\
d$_{rc}$ (pc) & --- &  7938 $\pm$ 78 &  7926 $\pm$ 87 & 68 & --- & --- & --- & --- & ---& ---\\
$\sigma_d$ (mag) & --- & 0.192 $\pm$ 0.004 & 0.187 $\pm$ 0.005 & 1x10$^{-4}$  & --- & --- & --- & --- & ---& ---\\
A$_{F814W}$ / A$_{Ks}$ & 7.81 $\pm$ 1.5 & --- &  9.66 $\pm$ 0.32 & 0.33 & 5.09 & 8.14 &  7.30 & 8.98 & 5.86 & 9.71\\
A$_{y}$ / A$_{Ks}$ & 5.13 $\pm$ 0.91 & --- &  6.29 $\pm$ 0.19 & 0.24  & 3.72 & 5.87 & 5.05 & 6.31 & 4.16 & 6.71 \\
A$_{F125W}$ / A$_{Ks}$ & 3.01 $\pm$ 0.43 & --- &   3.56 $\pm$ 0.10 & 0.11 & 2.44 & 3.42 & 2.93 & 3.57 & 2.56 & 3.81\\
A$_{F160W}$ / A$_{Ks}$ & 2.05 $\pm$ 0.23 & --- &  2.33 $\pm$ 0.06 & 0.04 &1.76 & 2.17 & 1.95 & 2.24 & 1.75 & 2.39 \\
A$_{[3.6]}$ / A$_{Ks}$ & 0.50 $\pm$ 0.03 & --- & 0.50 $\pm$ 0.03 & 0.002 & --- &  0.36 & 0.50 & --- & 0.54  & 0.18 \\
A$_{F125W}$ / A$_{F160W}$ &  1.468 $\pm$ 0.047 &  1.527 $\pm$ 0.006 & 1.525 $\pm$ 0.004 & 0.01 & 1.385 & 1.576 & 1.498 & 1.595 & 1.459 & 1.616 \\
$\Delta$ZP$_{Ks}$ (mag) & 0.0024 $\pm$ 0.008 & --- &  0.0025 $\pm$ 0.008 & 0.00014 & --- & --- & ---& --- & ---& ---\\
$\Delta$ZP$_{F160W}$ (mag) & -0.0037 $\pm$ 0.006 & ---  & -0.005 $\pm$ 0.006 & 0.0024  & --- & --- & ---& --- & ---& ---\\
$\Delta$ZP$_{F153M}$ (mag) & --- &  0.0013 $\pm$ 0.007 & 0.002 $\pm$ 0.006 & 0.0057 & --- & --- & ---& --- & ---& ---\\
$\Delta$ZP$_{F127M}$ (mag) & --- &  -0.0011 $\pm$ 0.007 & -0.0012 $\pm$ 0.007 & 0.0031 & --- & --- & ---& --- & ---& ---\\
$\Delta$ZP$_{F125W}$ (mag) & 0.0016 $\pm$ 0.006  & --- & 0.0022 $\pm$ 0.006 & 0.0014 & --- & --- & ---& --- & ---& ---\\
$\Delta$ZP$_{F1814W}$ (mag) & 0.0001 $\pm$ 0.0047 &  --- &  0.000 $\pm$ 0.007 & 0.00042 & --- & --- & ---& --- & ---& ---\\
\cutinhead{Derived Parameters}
$\mathcal{S}_{1/0.806 \mu m}$ & 3.49 $\pm$ 0.26  & --- & 3.50 $\pm$ 0.28 & 0.18 & 15.79 & 16.52 & 20.14  & 18.58 & 19.48 & 19.48\\
$\mathcal{S}_{1/0.962 \mu m}$ & 2.82 $\pm$ 0.16 & --- & 2.84 $\pm$ 0.16 &  0.12 & 8.03  & 10.95 & 11.67  & 11.96 & 13.52 & 13.52 \\
$\mathcal{S}_{1/1.25 \mu m}$ &  1.64 $\pm$ 0.09  & --- & 1.66 $\pm$ 0.08 & 0.04 & 4.06 & 5.52 & 5.00 & 5.66 & 5.65 & 5.65\\
$\mathcal{S}_{1/1.53 \mu m}$ &  1.54 $\pm$ 0.10  &  --- & 1.56 $\pm$ 0.10 & 0.07  & 2.39 & 3.04 & 2.73 & 3.01 & 3.40 & 3.40\\
$\mathcal{S}_{1/2.14 \mu m}$ & 1.0 & --- & 1.0 & 0 & 1.0 & 1.0 & 1.0 & 1.0 & 1.0 & 1.0 \\
\enddata
\label{tab:results}
\tablenotetext{a}{Assuming a Wd1 cluster age of 5 Myr}
\tablenotetext{b}{C89: \citet{Cardelli:1989qf}; D16: \citet{Damineli:2016no}; N09: \citet{Nishiyama:2009fc}; F09: \citet{Fitzpatrick:2009ys}, with $\alpha$=2.5; S16$_1$: \citet{Schlafly:2016cr}, rhk = 1.55; S16$_2$: \citet{Schlafly:2016cr}, rhk = 2.0}
\tablenotetext{c}{Systematic error, see $\mathsection$\ref{sec:sysError}}
\tablenotetext{d}{From \citet{Andersen:2017aq} using N09 law}
\end{deluxetable*}

\subsection{Wd1 MS Only}
\label{sec:Wd1results}
\begin{figure*}
\begin{center}
\includegraphics[scale=0.35]{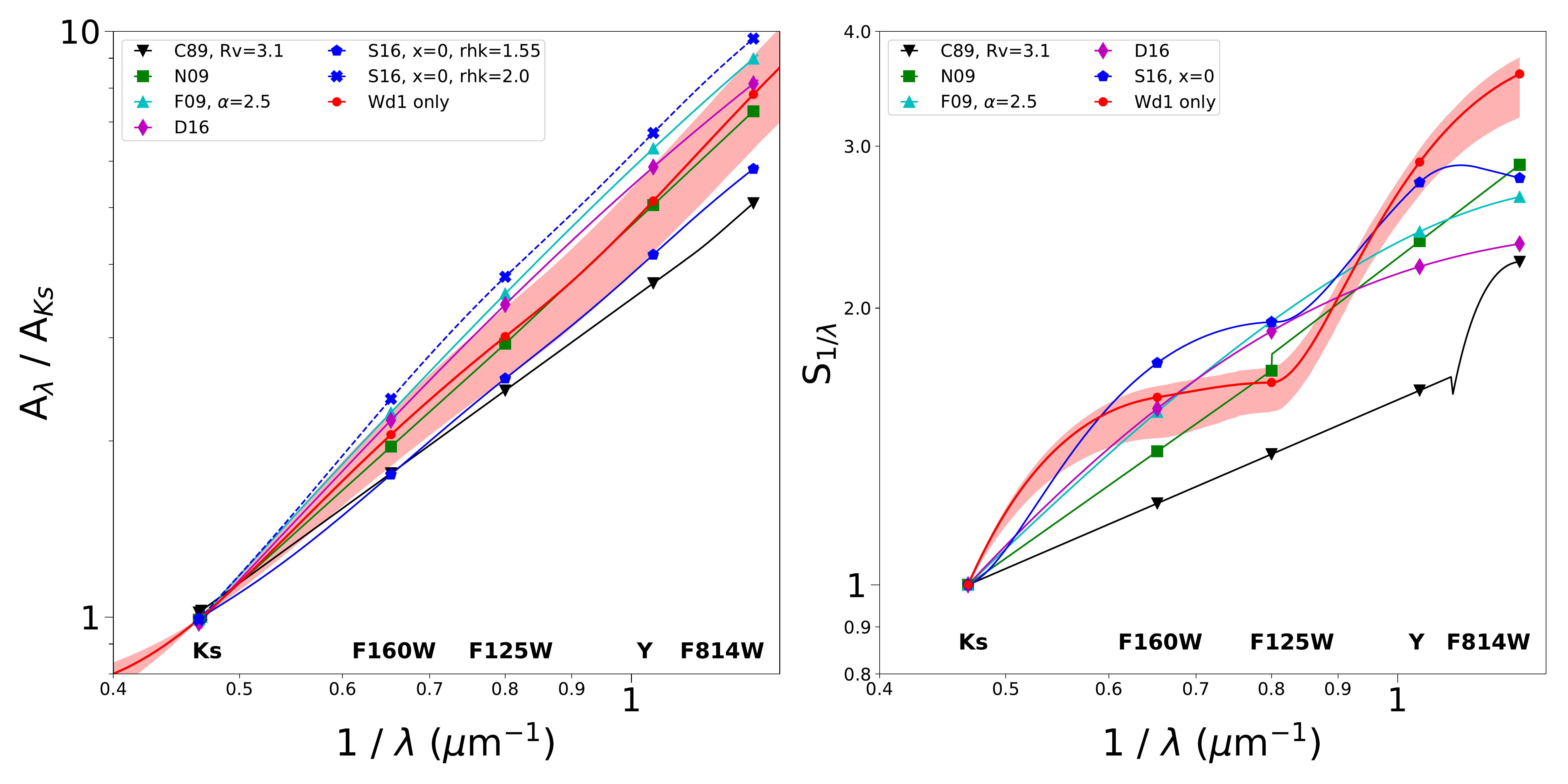}
\end{center}
\caption{
Extinction law fit to the Wd1-only sample, with the extinction law to the left and the shape of the law (in terms of $\mathcal{S}_{1/\lambda}$, see Equation \ref{eq:shape}) to the right. The best-fit model is represented by the solid red line, with the 1$\sigma$ statistical errors represented by the red shaded region. Several extinction laws from the literature are included for comparison: \citet[][C89]{Cardelli:1989qf}, \citet[][N09]{Nishiyama:2009fc}, \citet[][D16]{Damineli:2016no}, \citet[][S16]{Schlafly:2016cr} and \citet[][DM16]{De-Marchi:2016qw}. A cluster age of 5 Myr is assumed.
\label{fig:Wd1_deg}
}
\end{figure*}

\begin{figure*}
\begin{center}
\includegraphics[scale=0.27]{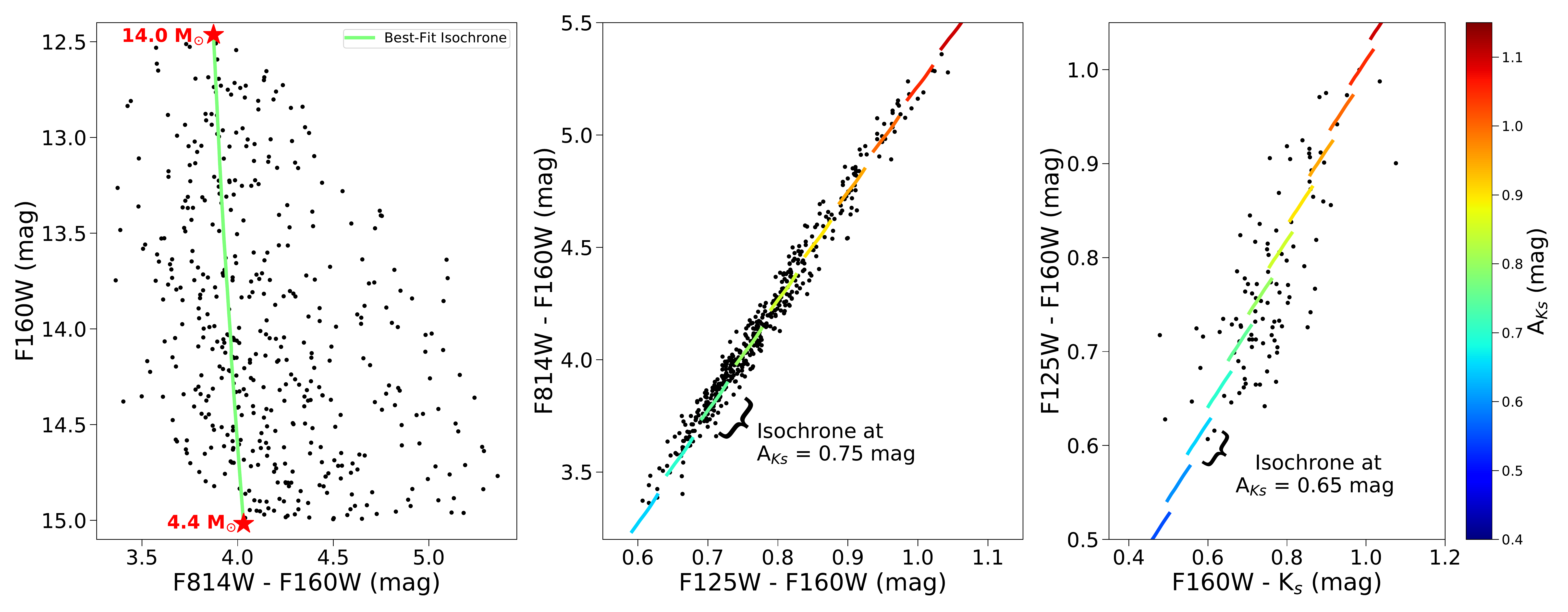}
\end{center}
\caption{
A comparison between the model isochrones and the observed Wd1 sample for the best-fit extinction law in the Wd1-only analysis. In each plot, the observed sample is shown by the black points and the model isochrones are shown as lines with a color corresponding to their total extinction. The left plot shows the best-fit isochrone in the CMD (A$_{Ks}$ = 0.78 mag) with the highest and lowest masses labeled by red stars. The middle and right plot show individual isochrones at different total extinctions in the \emph{HST} and \emph{HST}-\emph{VISTA} 2CD, respectively. These isochrones trace the reddening vector of the population.
\label{fig:Wd1_data}
}
\end{figure*}

To fit the Wd1 MS sample, we set $\mathcal{L}_{tot} =  \mathcal{L}_{wd1}$ in Equation \ref{eq:bayes} and do not consider the model parameters related to the RC stars (d$_{rc}$, $\sigma_{rc}$, ZP$_{F127M}$, and ZP$_{F153M}$). A cluster age of 5 Myr is assumed. As with the simulated cluster tests in Appendix \ref{app:art}, a large degeneracy is found between d$_{wd1}$, A$_{K_s}$ and the extinction law due to the uncertainty in the normalization, though the shape of the law is well constrained (Figure \ref{fig:Wd1_deg}). d$_{wd1}$ is constrained to 4428 $\pm$ 309 $\pm$ 48 pc, consistent with the result from \citep{Kothes:2007jx}, and A$_{K_s}$ is found to be 0.78 $\pm$ 0.16 $\pm$ 0.02 mag, also broadly consistent with the literature. No statistically significant zeropoint offsets are obtained for any of the filters, indicating that the zeropoints presented in in Figure \ref{fig:ZP} are accurate to 0.006 mag for the \emph{HST} filters and 0.008 mag for the \emph{VISTA} K$_s$ filter. The extinction law provides a good fit to the observations, with reduced chi-square ($\chi^2_{red}$) values of 0.38 for the \emph{HST} CCD and 0.77 for the \emph{HST}-\emph{VISTA} CCD (Figure \ref{fig:Wd1_data}). That the $\chi^2_{red}$ values are less than 1.0 suggests that the photometric errors are conservative.

The shape of the extinction law is inconsistent with a single power law, which would produce a straight line in S$_{1/\lambda}$ vs. $\log(1/\lambda)$ plot in Figure \ref{fig:Wd1_deg}. We will examine this further in the final extinction law fit ($\mathsection$\ref{sec:Wd1Archesresults}).

\begin{figure*}
\begin{center}
\includegraphics[scale=0.35]{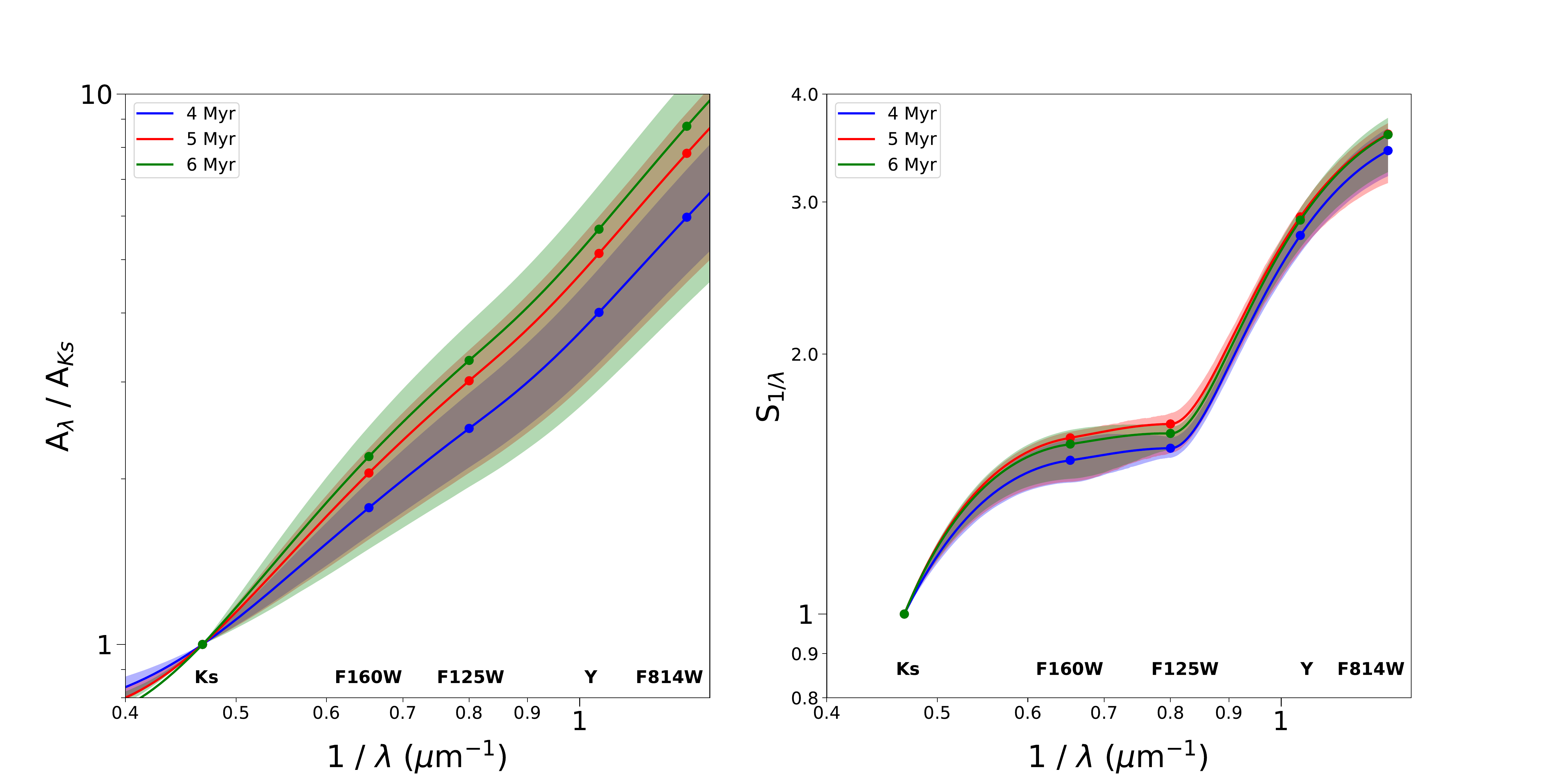}
\end{center}
\caption{
Extinction law fits to the Wd1 sample assuming cluster ages of 4 Myr (blue), 5 Myr (red), and 6 Myr (green). The best-fit model is represented by the solid lines, with the 1$\sigma$ statistical limits represented by the shaded regions. While the extinction law degeneracy increases when different ages are considered (left plot), the extinction law shape is effectively unchanged (right plot).
\label{fig:age_deg}
}
\end{figure*}

We test the impact of assuming an age for Wd1 by performing identical analyses with cluster ages of 4 Myr and 6 Myr, the range of possible ages suggested by the evolved star population \citep{Crowther:2006hb, Negueruela:2010hc}. Since the stellar mass of the pre-MS bridge decreases with age, changing the age affects the mass range of the MS sample. Using the selection criteria described in $\mathsection$\ref{sec:Iso}, the MS mass range becomes 4.83~M$_{\odot}$ -- 16.0~M$_{\odot}$ for a 4 Myr cluster and 3.95~M$_{\odot}$ -- 12.50~M$_{\odot}$ for a 6 Myr cluster. This affects the normalization of the extinction law, with the law generally becoming steeper (i.e., larger A$_{\lambda}$~/~A$_{Ks}$ values) with increasing age, but the shape of the law remains almost identical (Figure \ref{fig:age_deg}).

\subsection{Arches RC stars only}
\label{sec:Archesresults}
Since the distance distribution of the Arches RC population is known to significantly higher precision than Wd1, the normalization of the extinction law can be much better constrained. We perform an extinction law fit using just the RC sample, setting $\mathcal{L}_{tot} =  \mathcal{L}_{rc}$ in Equation \ref{eq:bayes} and removing the free parameters related to Wd1 (i.e. d$_{wd1}$, A$_{K_s}$). Only the extinction law ratios A$_{F125W}$~/~A$_{K_s}$ and A$_{F160W}$~/~A$_{K_s}$ are included in the model, approximately matching the wavelengths of the F127M and F153M observations. We present the ratio A$_{F125W}$~/~A$_{F160W}$, which combines the information from both filters.

The best-fit recovers A$_{F125W}$~/~A$_{F160W}$ = 1.527 $\pm$ 0.006 $\pm$ 0.01 and d$_{rc}$ = 7938 $\pm$ 87 $\pm$ 68 pc.  The statistical uncertainty in A$_{F125W}$~/~A$_{F160W}$ is a factor of $\sim$8 smaller than the value obtained for the Wd1-only fit (1.468 $\pm$ 0.047 $\pm$ 0.01), eliminating much of the uncertainty in the normalization. Statistically, d$_{rc}$ is constrained to $\sim$1.1\%, which is better than the input prior and in principle places a constraint on the GC distance. However, we have adopted a simplified description of the RC (assuming an average age and metallicity for the population) and are susceptible to possible systematic uncertainties in the RC star model ($\mathsection$\ref{sec:sysRC}). A careful analysis of the GC distance is beyond the scope of this paper.

\subsection{Wd1 MS and Arches RC Combined Fit}
\label{sec:Wd1Archesresults}
\begin{figure*}
\begin{center}
\includegraphics[scale=0.35]{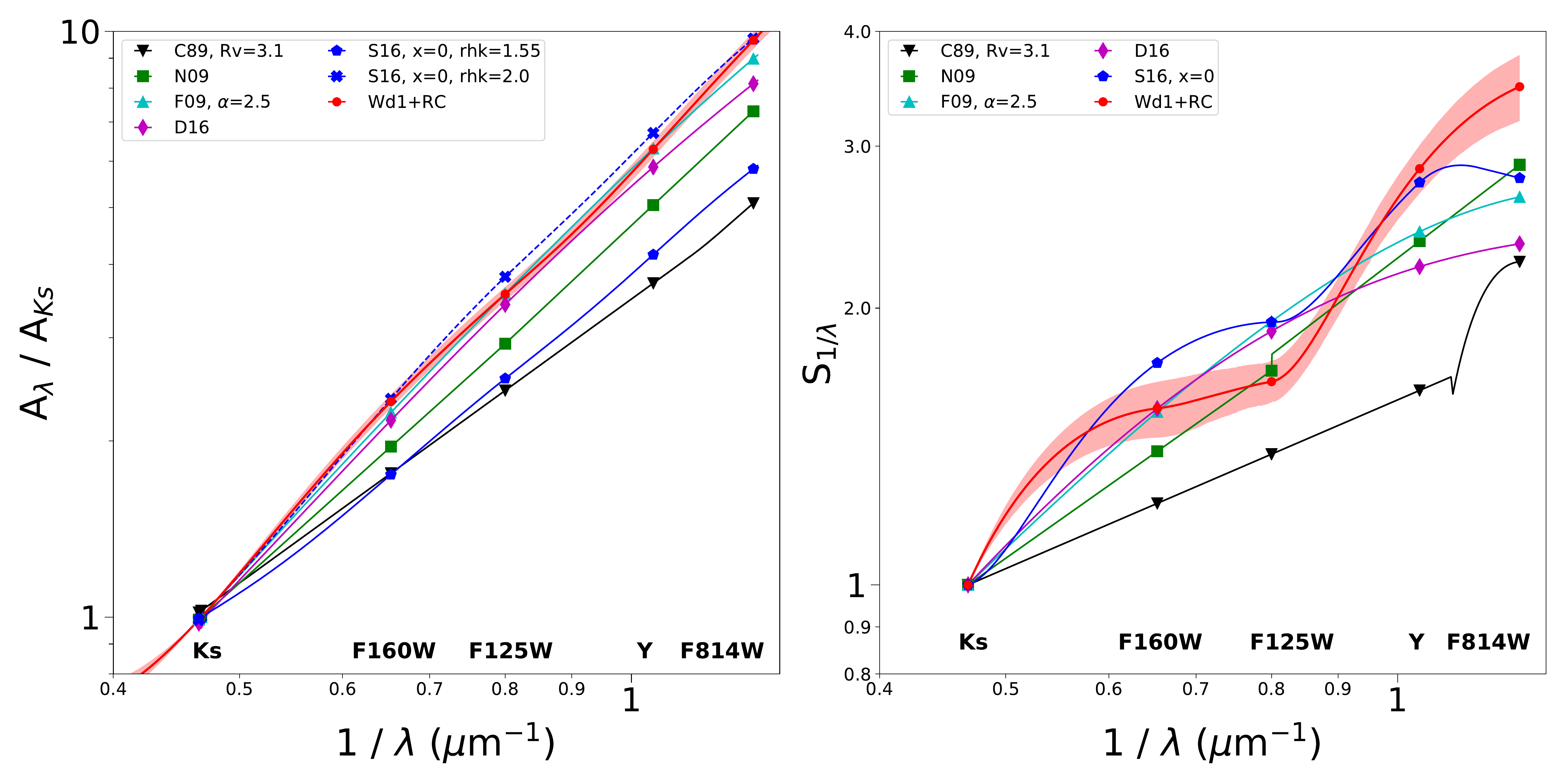}
\end{center}
\caption{
Extinction law fit to the Wd1+RC sample, with the extinction law to the left and the shape of the law (in terms of $\mathcal{S}_{1/\lambda}$) to the right. The best-fit model is represented by the solid red line, with the 1$\sigma$ errors represented by the red shaded region. The observed law is inconsistent with a power law (which would appear as a straight line in both plots) and is most similar to the F09 law (with $\alpha$ = 2.5) and S16 law (with rhk = 2.0) from the literature. A Wd1 age of 5 Myr is assumed, though this assumption has no significant effect on either the law or S$_{1/\lambda}$.
\label{fig:wd1_rc_law}
}
\end{figure*}

\begin{figure*}
\begin{center}
\includegraphics[scale=0.30]{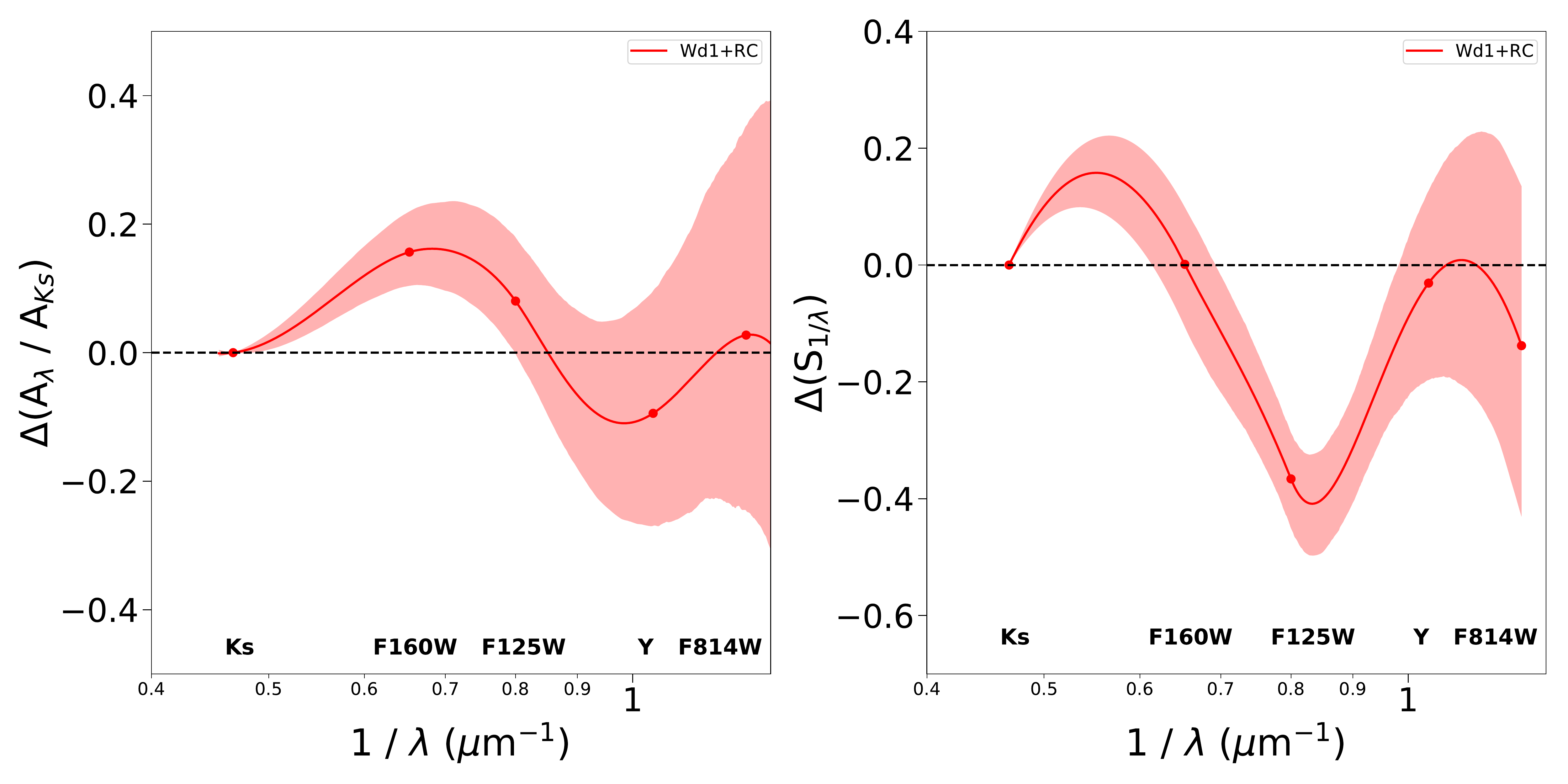}
\end{center}
\caption{
The deviation of the Wd1+RC extinction law from a power law, both in terms of A$_{\lambda}$~/~A$_{Ks}$ (left) and S$_{1/\lambda}$ (right). The residuals between the best-fit law and a power law are shown by the red points and lines, while the uncertainties are represented by the red shaded regions. The Wd1+RC extinction law is statistically inconsistent with a power law, even in the NIR.
\label{fig:plComp}
}
\end{figure*}

The final extinction law derived using the combined Wd1 MS + Arches RC sample is shown in Figure \ref{fig:wd1_rc_law}. Once again, a Wd1 age of 5 Myr is assumed. The law is well constrained with combined uncertainties (statistical and systematic added in quadrature) better than $\sim$5\% in A$_{\lambda}$~/~A$_{K_s}$ and $\sim$10\% in $\mathcal{S}_{1/\lambda}$ (Table \ref{tab:results}). There is no significant change in the shape of the law relative to the Wd1-only fit, which supports the assumption that the shape is the same for both populations. Fitting a power law to the A$_{\lambda}$~/A$_{Ks}$ values results in a reduced chi-squared ($\chi^2_{red}$) value of 3.7, indicating that the difference between the best-fit law and a power law are statistically significant (Figure \ref{fig:plComp}). This is in agreement with previous studies of the OIR extinction law \citep[e.g.][]{Fitzpatrick:2009ys}. Interestingly, we find that the NIR portion of the law (K$_s$, F160W, F125W) is also statistically inconsistent with a power law with $\chi^2_{red}$ = 3.3, in contrast to what is often assumed in the literature. We only consider the statistical errors on A$_{\lambda}$~/~A$_{Ks}$ in the $\chi^2_{red}$ calculations. Extinction law analyses with various systematics applied ($\mathsection$\ref{sec:sysError}) also show statistically significant deviations from a power law, and so the systematics do not affect this conclusion. A stand-alone python code to calculate the Wd1+RC extinction law at all wavelengths between 0.8 $\mu$m -- 2.2 $\mu$m is available online\footnote{https://doi.org/10.5281/zenodo.1063708}.

\begin{figure*}
\begin{center}
\includegraphics[scale=0.27]{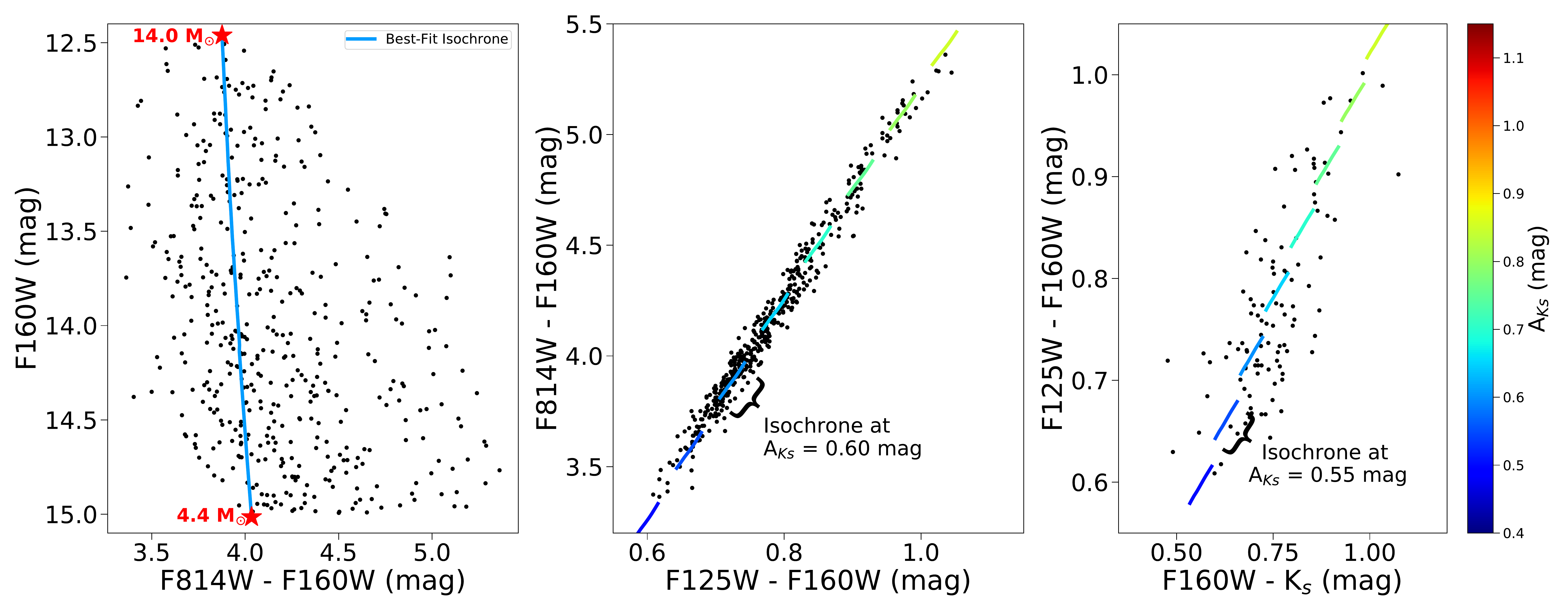}
\end{center}
\hspace{2.27in} \includegraphics[scale=0.27]{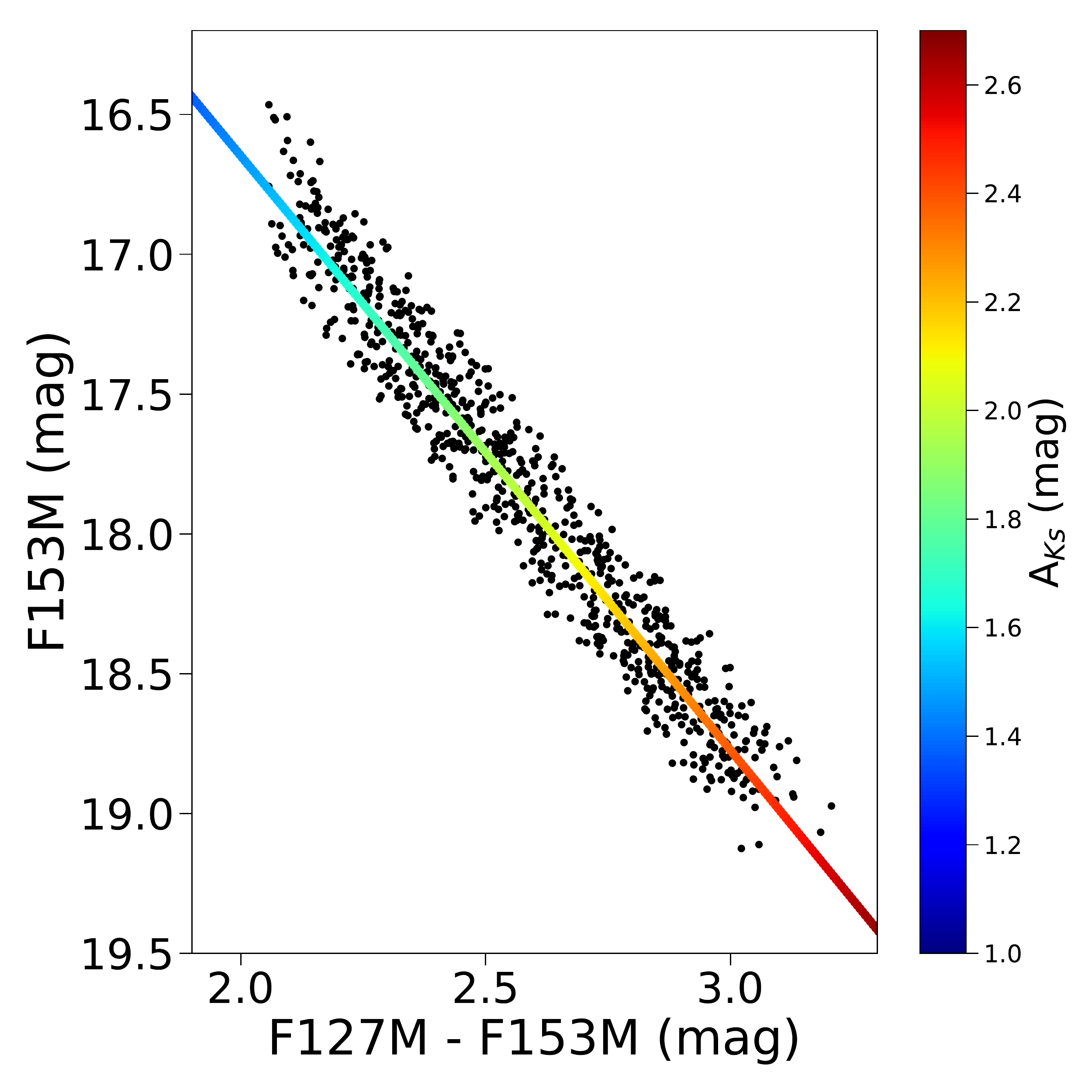}
\caption{
A comparison between the model isochrones and the data for the best-fit extinction law in the Wd1+RC analysis. \emph{Top:} The Wd1 sample and cluster isochrones plotted in CMD space (left) and the \emph{HST} and \emph{HST - VISTA} 2CD (middle and right, respectively), using the same conventions as Figure \ref{fig:Wd1_data}. \emph{Bottom:} The Arches RC stars (black points) compared to the best-fit reddening vector (colored line, where the color at each point corresponds to the total extinction) in the CMD. The model provides a good match to the data for both the Wd1 and RC samples.
\label{fig:wd1_rc_data}
}
\end{figure*}

A comparison of the best-fit model and Wd1 and Arches RC observations is shown in Figure \ref{fig:wd1_rc_data}. The law is able to reproduce the photometry of both populations well, with $\chi^2_{red}$ values for the Wd1 CCDs are nearly identical to the Wd1-only fit (0.38 and 0.77, respectively) and $\chi^2_{red}$ = 0.83 for the RC CMD.

Repeated analyses of the extinction law assuming Wd1 ages of 4 Myr, 6 Myr, and 7 Myr show that the extinction law is not impacted by changing the cluster age. However, d$_{wd1}$ is found to systematically decrease with increasing cluster age. We explore this trend further in $\mathsection$\ref{sec:ageDistance}.

\subsection{The Age and Distance of Wd1}
\label{sec:ageDistance}
To demonstrate of the impact of the extinction law, we show how the new extinction law changes the distance and age of the cluster. In the Wd1+RC extinction law fit, d$_{wd1}$ varies from 5222 $\pm$ 113 pc for a Wd1 age of 4 Myr to 4133 $\pm$ 66 pc for a Wd1 age of 7 Myr (statistical and systematic errors added in quadrature). Thus, an independent estimate of d$_{wd1}$ offers a constraint on the cluster's age. This is especially valuable given the diverse population of evolved stars in Wd1, which include yellow hypergiants \citep{Clark:2005sp}, red supergiants \citep{Clark:2010vn}, WR stars \citep{Crowther:2006hb}, luminous blue variables \citep{Clark:2004ly, Dougherty:2010gf}, and a magnetar \citep{Muno:2006zr}. The presence of these objects provides a strong test of stellar evolution models, which struggle to reproduce such a sizable population of cool supergiants and WR stars simultaneously \citep{Clark:2010vn, Ritchie:2010eq}.

We obtain an independent estimate of d$_{wd1}$ from published measurements of W13, a 9.2 day eclipsing binary within the cluster \citep{Bonanos:2007nh}. \citet[][hereafter K12]{Koumpia:2012ij} combine the optical (VRI) lightcurves from \citet{Bonanos:2007nh} with multi-epoch spectroscopy to derive the physical properties of the system, found to be a near-contact binary composed of a B0.5Ia+/WNVL and O9.5-B0.5I star. From the derived effective temperatures  and stellar radii, they calculate a total system luminosity of $\log{L / L_{\odot}}$ = 5.54 $\pm$ 0.11. K12 correct for extinction by adopting A$_J$~/~A$_{Ks}$ = 2.50 $\pm$ 0.15 \citep{2005ApJ...619..931I}, ultimately reporting a distance of 3710 $\pm$ 550 pc. We redo this calculation using the Wd1+RC extinction law, which is steeper than the Indebetouw value (A$_{J}$~/~A$_{Ks}$ = 3.56 $\pm$ 0.15, with statistical and systematic errors added in quadrature). This does not bias the Wd1 distance calculation since the law is independent of cluster age and thus not tied to the value of d$_{wd1}$ derived in the extinction law analysis.

Following K12, we calculate a distance to W13 using its 2MASS J-band apparent magnitude, a theoretical bolometric correction (BC$_{\lambda}$) for O9.5I stars, and an extinction correction based on the NIR colors of nearby WR stars. The 2MASS J-band magnitude is 9.051 $\pm$ 0.16 mags, with the uncertainty set by the depth of the primary eclipse since the phase of the measurement is unknown. From \citet{Martins:2006qq} we adopt BC$_{J}$ = -3.24 $\pm$ 0.08 mags, with the uncertainty based on the scatter in the BC$_{\lambda}$ - T$_{eff}$ relation. The total extinction (A$_{Ks}$) of W13 is calculated from the J~-~K color excesses of the Wolf-Rayet stars ``R" and ``U" reported in \citet{Crowther:2006hb}, both of which are $\sim$6" away from W13. Using our extinction law, these stars have A$_{Ks}$ = 0.55 $\pm$ 0.05 mags and 0.52 $\pm$ 0.05 mags, respectively, and so we adopt A$_{Ks}$ = 0.535 $\pm$ 0.035 mag for W13.

With all the pieces in place, we can calculate a distance to W13:

\begin{equation}
\mu = m_J - M_J - A_J
\end{equation}

\noindent where $\mu$ is the distance modulus, m$_J$ is the 2MASS J-band apparent magnitude, A$_J$ is the total extinction at J-band, and M$_J$ is the J-band absolute magnitude. The J-band absolute magnitude is calculated from the luminosity using the bolometric correction:

\begin{equation}
M_J = M_{\odot}^{bol} - BC_J - 2.5 \log{\left(\frac{L}{L_{\odot}}\right)}
\end{equation}

\noindent where M$_{\odot}^{bol}$ = 4.75 is the bolometric magnitude of the sun and L$_{\odot}$ is the solar luminosity. Plugging in the values and propagating the errors, we find $\mu$ = 12.958 $\pm$ 0.235 mag, or 3905 $\pm$ 422 pc. The major source of remaining uncertainty is the 2MASS photometry, since the phase of the system was unknown at the time of measurement. New multi-epoch OIR photometry of W13 would dramatically increase the precision of the derived distance. Unfortunately W13 is saturated in the \emph{HST} observations and so our observations are not helpful in this regard.

\begin{figure}
\begin{center}
\includegraphics[scale=0.35]{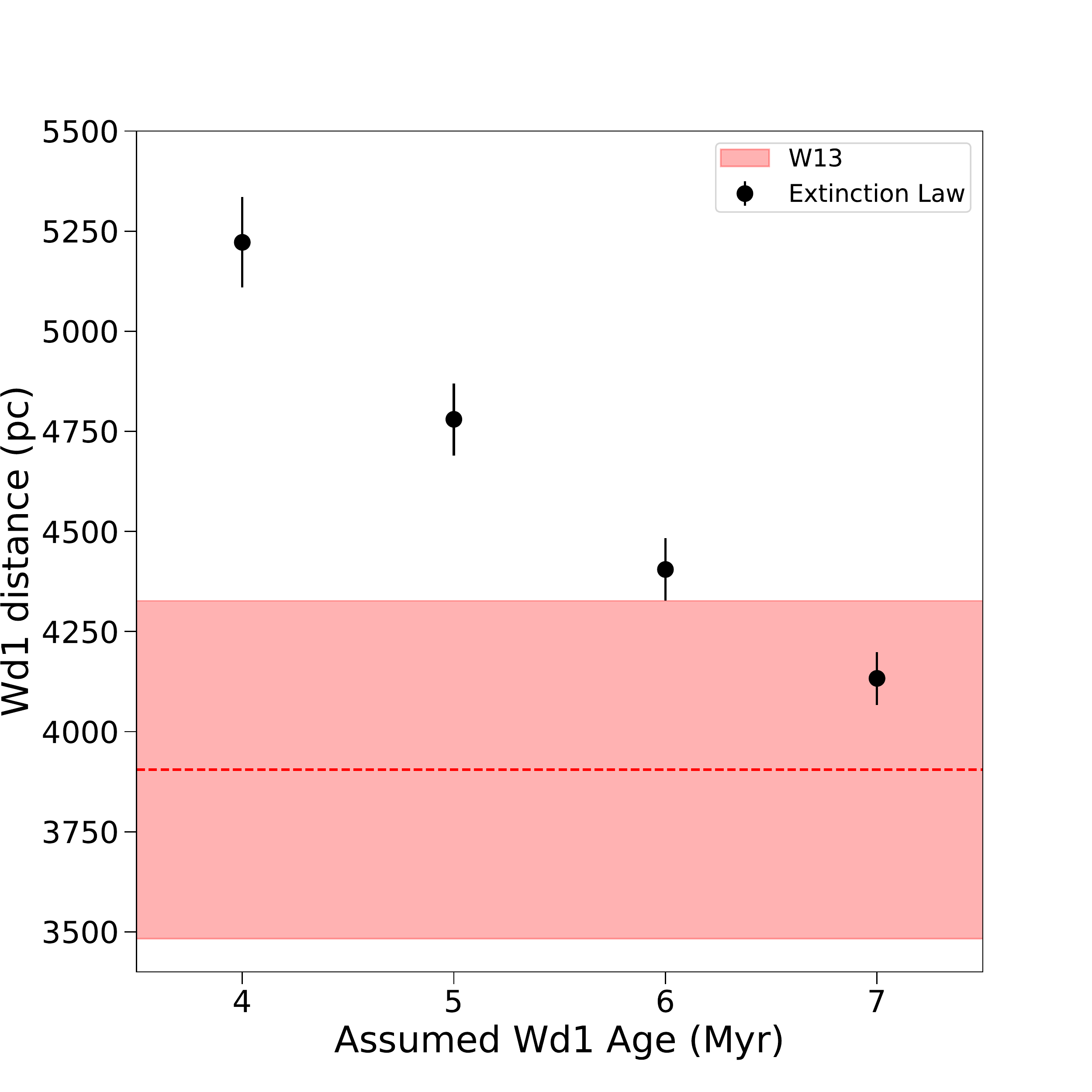}
\end{center}
\caption{
Wd1 distance as a function of the assumed cluster age in the Wd1+RC extinction law analysis. No other parameter shows statistically significant variations. The red dotted line shows the independent distance estimate from the eclipsing binary W13 using the best-fit law, with the 1$\sigma$ error represented by the red shaded region. A cluster age of 6 -- 7 Myr is favored over an age of 4 -- 5 Myr.
\label{fig:EBdist}
}
\end{figure}

A comparison between the W13 distance and the d$_{wd1}$ from the extinction law analysis is shown in Figure \ref{fig:EBdist}. The eclipsing binary distance is consistent with an older cluster age, being discrepant from the 4 Myr and 5 Myr extinction fit distances by 3.0$\sigma$ and 2.0$\sigma$, respectively. The 6 Myr and 7 Myr distances are only discrepant by 1.2$\sigma$ and 0.5$\sigma$, respectively. An older cluster age for Wd1 is not unreasonable, given that the constraint of 4 -- 5 Myr from \citet{Crowther:2006hb} is based stellar evolution models of the Wolf-Rayet star population, which are not yet well understood. \citet[][]{Negueruela:2010hc} find that the HR diagram of OB supergiants is consistent with a cluster age greater than 5 Myr, though this analysis relies on the \cite{Rieke:1985dw} extinction law (A$_J$~/A$_{Ks}$ = 2.35). Unfortunately their observations are primarily shortward of I-band, and so we cannot recreate their HR diagram using our extinction law. We leave a detailed analysis of the age and distance of Wd1 to a future paper.

\section{Discussion}
\subsection{Comparison with Previous Extinction Laws}
\label{sec:comparison_discussion}
We compare our result with several extinction laws in the literature. These include \citet[][hereafter C89]{Cardelli:1989qf}, assuming R$_V$ = 3.1 as is often adopted for the interstellar medium (however, changing R$_V$ has little effect on the extinction law in this wavelength range); \citet[][hereafter N09]{Nishiyama:2009fc}, often used for the GC and Galactic bulge; \citet[][hereafter D16]{Damineli:2016no}, derived specifically for Wd1; \citet[][hereafter F09]{Fitzpatrick:2009ys}, assuming $\alpha$ = 2.5 and R$_V$ = 3.1; and \citet[][hereafter S16]{Schlafly:2016cr}, assuming A$_H$~/~A$_{Ks}$ (``rhk'') values of 1.55 (consistent with \citeauthor{2005ApJ...619..931I} 2005) and 2.0. To construct the C89, D16, and F09 laws we use the functional forms reported by the authors, and for S16 we use the python code referenced in their Appendix\footnote{http://faun.rc.fas.harvard.edu/eschlafly/apored/extcurve\_s16.py}. For N09 we adopt a power law with $\beta$ = 2.0 between 1.25 $\mu$m -- 2.14 $\mu$m and then use a linear interpolation in log(A$_{\lambda}$~/~A$_{Ks}$) vs. log(1 / $\lambda$) space to extend from A$_{J}$~/~A$_{Ks}$ to the A$_{V}$~/~A$_{Ks}$ value reported in \citet{Nishiyama:2008wa}. This method was adopted to have minimal impact on the shape of the N09 law shortward of J-band, where no functional form is provided.

\begin{figure*}
\begin{center}
\includegraphics[scale=0.35]{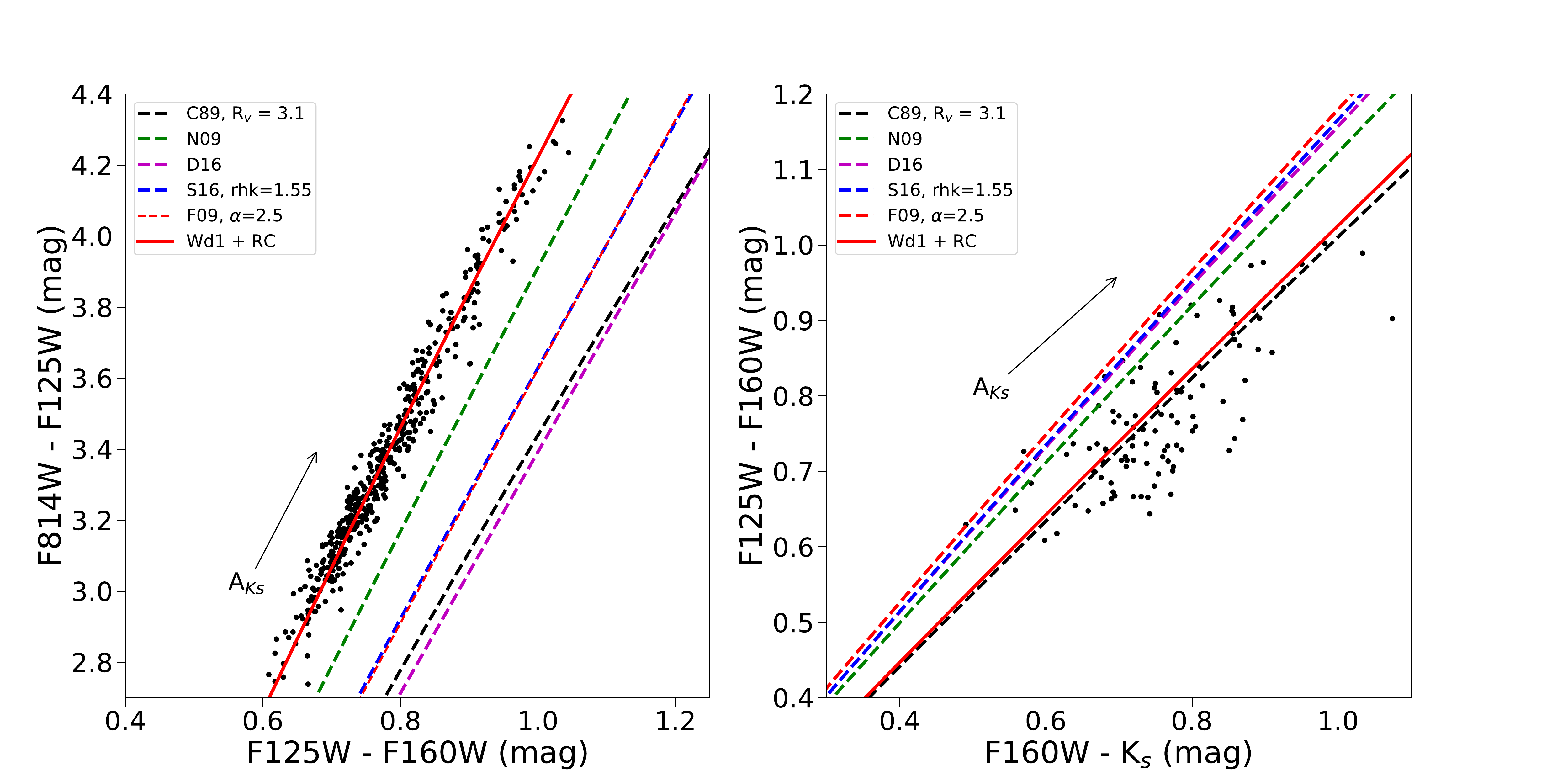}
\end{center}
\caption{
Color-color diagrams comparing the observed Wd1 MS to the reddening vectors predicted by Wd1 + RC extinction law (red solid line) and several extinction laws in the literature (dashed lines). The literature laws show significant differences between the reddening vector and observations, especially in the \emph{HST} colors where the photometric errors are typically 0.02 mag. These discrepancies are caused by differences in the extinction law shape. Note that all reddening vectors trace back to the un-reddened main sequence (approximately at the origin of these plots), but diverge at the high extinction of Wd1 due to varying degrees of curvature in the vector.}
\label{fig:tcd_iso}
\end{figure*}

Theoretical cluster isochrones using the published extinction laws are unable to reproduce the observed colors of the Wd1 MS (Figure \ref{fig:tcd_iso}). This is especially evident in the \emph{HST} 2CD, where the model reddening vectors are offset between 0.29 mag -- 0.72 mag to the blue in F814W - F125W (about 9\% and 22\%, respectively), and between 0.08 mag -- 0.19 mag to the red in F125W - F160W (about 12\% and 25\%, respectively). These offsets are huge relative to the \emph{HST} photometric errors, which is typically 0.02 mag for both colors. In the \emph{HST}-\emph{VISTA} 2CD, most literature extinction laws produce reddening vectors that are 0.07 mag -- 0.10 mag too blue in F160W - K$_{s}$ (about 12\%), which is significant but not as large relative to the typical color uncertainty of $\sim$0.06 mags. The exception is C89, which reproduces the observed sequence in this 2CD well. These isochrones assume a cluster age of 5 Myr, though changing the age has little to no effect on the colors. Note that the discrepancies in color-color space are due to differences in the shape of the extinction law, rather than the normalization.

In Figures \ref{fig:Wd1_deg}, \ref{fig:wd1_rc_law} and Table \ref{tab:results} we compare our extinction law results to the published laws. The Wd1+RC extinction law is generally steep (i.e. has large A$_{\lambda}$~/~A$_{Ks}$ values), being most similar to F09 law with $\alpha$ = 2.5 and the S16 law with A$_H$ / A$_{Ks}$ = 2.0. Notably, we find A$_{F125W}$~/~A$_{Ks}$ to be 18\% larger and A$_{F814W}$~/~A$_{Ks}$ to be 24\% larger than the N09 law commonly used for the inner Milky Way. The shape of the law also differs from those in the literature, which is expected given the inability of the published laws to reproduce the observations in Figure \ref{fig:tcd_iso}. In particular, $S_{F814W}$~/~$S_{Ks}$ is significantly larger than any of the published laws, indicating a steeper derivative through the F814W filter relative to the K$_s$ filter. However, our data calls for such steepness as discussed in $\mathsection$\ref{sec:sensitivity}.

Our extinction law differs from the one previously derived for Wd1 by D16, who describe the law as a power law with a slope $\beta$ = 2.13 $\pm$ 0.08 between 0.8 $\mu$m -- 4.0 $\mu$m. Our Wd1+RC law is 4\% larger at A$_{F125W}$~/~A$_{Ks}$ and 16\% larger at A$_{F814W}$~/~A$_{Ks}$. The D16 law is derived from ground-based photometry of 105 evolved stars and a sample of RC stars in the Wd1 field, using color excess ratios and requiring a power law functional form between the JHK$_s$ filters. We believe our study offers several key advantages: 1) space-based photometry with higher precision ($\sim$0.01 mag) than is generally possible for ground-based observations; 2) a sample of kinematically-selected Wd1 members that is $\sim$4x larger than the D16 sample and is composed of main sequence stars, where stellar models are better understood relative to evolved stars; and 3) our forward-modeling technique makes no assumption regarding the functional form of the extinction law. However, D16 does have the advantage of using RC stars in the Wd1 field, while we rely on RC stars toward the Arches cluster. Unfortunately, the \emph{HST} data does not stretch as far to the red as the D16 observations, and so we do not observe a significant RC population in the Wd1 field. That said, only our law is able to reproduce the observed MS colors of the cluster, indicating its effectiveness.

\subsection{Extinction Law at the Galactic Center}
\label{sec:GC}
\citet[][hereafter S10]{Schodel:2010eq} measure the total extinction of the GC in the H and K$_s$ filters using observations of RC stars in the region. Adopting a K$_s$ absolute magnitude M$_{Ks}$ = -1.54 mag \citep{Groenewegen:2008ph}, an intrinsic H - K$_s$ color (H - K$_s$)$_{0}$ = 0.07 mag, and a GC distance of 8.03 kpc, they calculate absolute extinction values of A$_H$ = 4.48 $\pm$ 0.13 mag and A$_{Ks}$ = 2.54 $\pm$ 0.12 mag. This results an extinction ratio A$_H$~/~A$_{Ks}$ = 1.76 $\pm$ 0.18, providing an independent test of the normalization of the extinction law.

We calculate A$_{H}$~/~A$_{Ks}$ for the Wd1+RC extinction law as well as the N09 law, which is often adopted for the GC/Galactic Bulge. Calculating A$_{H}$ and A$_{Ks}$ at effective wavelengths of 1.677 $\mu$m and 2.168 $\mu$m, respectively, the Wd1+RC law produces A$_{H}$~/~A$_{Ks}$ = 1.936 $\pm$ 0.08 (statistical and systematic combined in quadrature). This is 10\% higher than the S10 result but within 1$\sigma$ given the uncertainties. The N09 law predicts a value of 1.68 $\pm$ 0.03, which is 5\% lower than the S10 value (0.4$\sigma$ difference). However, it is important to note that the A$_{H}$~/~A$_{Ks}$ value in S10 and the extinction laws from this work and N09 rely on knowing the absolute magnitude, colors, and distance of RC stars in the filters of interest. This can lead to systematic errors in the extinction law, as discussed in $\mathsection$\ref{sec:sysError}. We conclude that our law is in acceptable agreement with the S10 measurement.

The Wd1+RC law is also broadly consistent with the most recent measurement of the GC NIR extinction law, which reports a power law with $\beta$ = 2.31 $\pm$ 0.03 \citep{2017arXiv170909094N}. A power-law fit to the NIR filters in our law results in $\beta$~=~2.38 $\pm$ 0.15. However, we reiterate our result in $\mathsection$\ref{sec:Wd1Archesresults} that the Wd1+RC law is inconsistent with a power law in the NIR regime. Non-power law behavior is hinted in \citet{2017arXiv170909094N} as small differences between power law exponents derived in the JH vs. HK$_s$ filters, though they conclude a single power law is sufficient within the sensitivity of their study.

\subsection{Curvature in the Reddening Vector}
\label{sec:sensitivity}
By forward modeling the Wd1 photometry we can account for nonlinearity in the reddening vector. This can occur in regions of high extinction, where the extinction-weighted central wavelength of a filter (i.e., the flux-weighted average wavelength of the extinguished stellar energy distribution convolved with the filter function) changes relative to another filter. Differences in the filter width \citep[e.g.][]{Kim:2005jw, Kim:2006os} or slope of the extinction law between filters can cause this effect. Both processes are captured by the synthetic photometry in our extinction law analysis.

\begin{figure*}
\begin{center}
\includegraphics[scale=0.35]{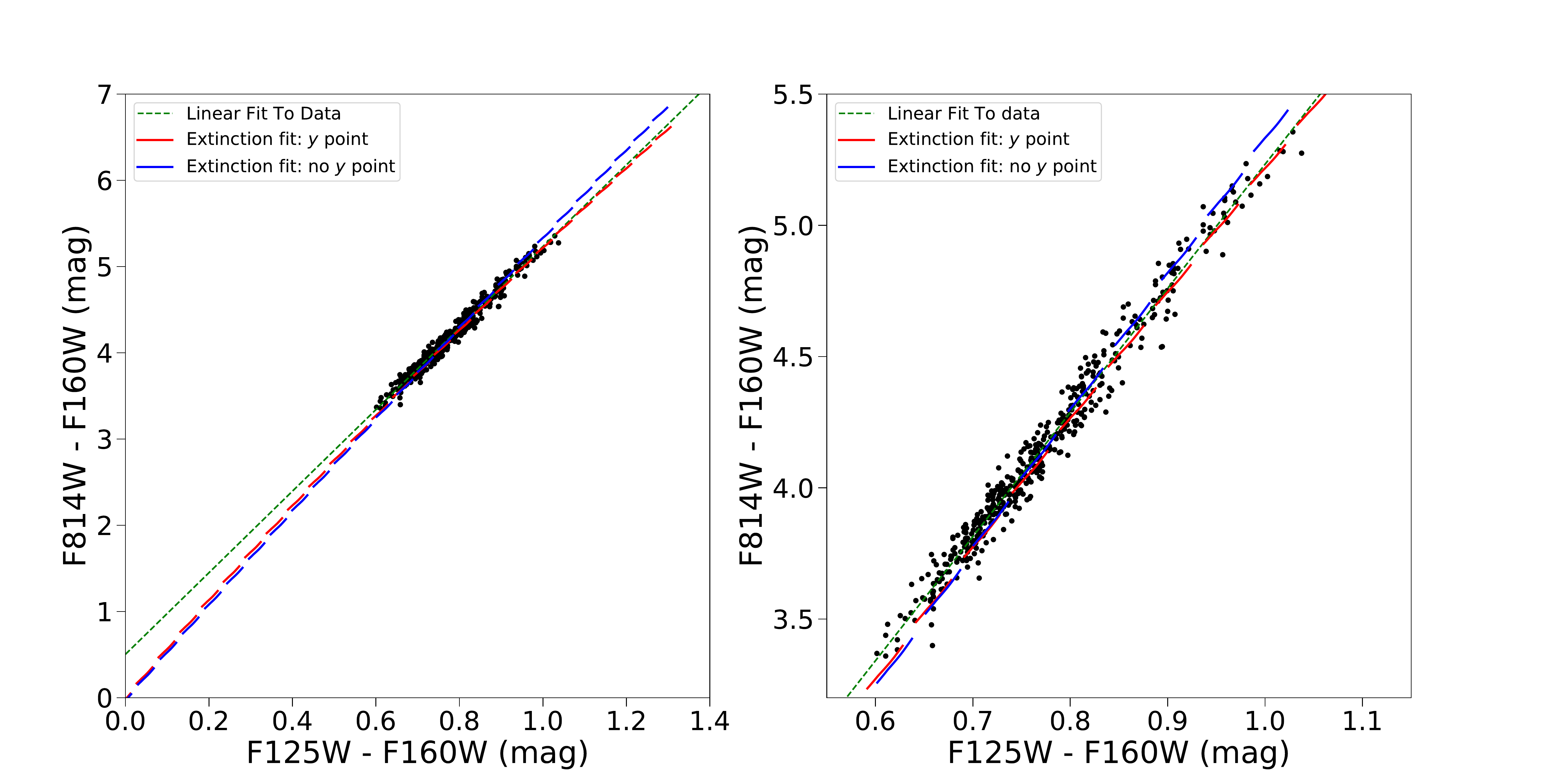}
\end{center}
\caption{
The \emph{HST} 2CD compared to an orthogonal linear regression fit to the data (green dashed line) and the extinction law models derived with (red) and without (blue) the PanStarrs $y$ point included. The right plot a zoomed-in version of the left plot. The fact that the linear fit does not trace back to the origin shows that there is significant curvature in the reddening vector. The model with the $y$ point, which allows for a steeper extinction law slope through the F814W filter, provides the best match to the data.
\label{fig:ypoint}
}
\end{figure*}

We observe reddening vector curvature in the \emph{HST} 2CD, where a simple linear fit to the MS sample does not trace back to the origin at A$_{Ks}$ = 0 as expected for these stars (Figure \ref{fig:ypoint}). Two extinction law fits to the data are shown: one with the PanStarrs $y$ point (0.962 $\mu$m) included and another without. Both vectors have significant curvature and are thus able to broadly match the data and trace back to the origin in the zero extinction case. However, the model with the $y$ point has more curvature and is able to better fit to the data. This is because adding A$_y$~/~A$_{Ks}$ increases the flexibility of the extinction curve between the F125W and F814W filters, allowing for a steeper slope through the F814W filter as is called for by the data.

\subsection{Sources of Systematic Error}
\label{sec:sysError}
Here we explore potential sources of systematic error and quantify their impact on the extinction law analysis.

\subsubsection{RC Star Model}
\label{sec:sysRC}
A possible source of systematic error is the RC stellar model, which is adopted from a 10 Gyr Parsec isochone at solar metallicity. The absolute F153M magnitude of the RC star model is used in combination with the GC distance to derive the overall extinction of the RC stellar population, which in turn sets the normalization of the Wd1+RC extinction law. If we use an intrinsic RC star model that is brighter than our current model, then the extinction law becomes systematically more shallow (i.e., smaller A$_{\lambda}$~/A$_{Ks}$ values), and vice versa.

Since there are no empirical calibrations of the RC absolute magnitude in either F127M or F153M, we must rely on the stellar models in our analysis. However, several measurements of the RC absolute magnitude exist for the 2MASS K$_s$ filter. The RC model we adopt is a 1.03 M$_{\odot}$ star with an absolute magnitude of K$_s$~=~-1.43 mag. An initial K$_{s}$ calibration from RC stars in the \emph{Hipparcos} catalog found an absolute magnitude of K$_s$ = -1.61 $\pm$ 0.03 mag \citep{Alves:2000it}, though this has been revised faintward to K$_{s}$ = -1.54 $\pm$ 0.04 mag \citep{Groenewegen:2008ph} and K$_s$ = -1.51 $\pm$ 0.01 \citep{2014MNRAS.441.1105F} based on updated \emph{Hipparcos} parallaxes \citep{van-Leeuwen:2007kh} and correcting for selection biases.

While the revised RC magnitudes are $\sim$0.1 mag brighter than our RC model, some or all of the discrepancy could be attributed to an age difference between the RC stars in our sample and those in the \emph{Hipparcos} sample. The Arches field RC stars are primarily located in the Galactic Bulge, which has been shown to be dominated by $\sim$10 Gyr stars \citep{Zoccali:2003dw, Clarkson:2008hw, Schultheis:2017zk}, though a spread of ages are present \citep[e.g.][]{Bensby:2013qt}. However, \emph{Hipparcos} sample reflects the star formation history of the local solar neighborhood. Studies of the solar neighborhood have suggested a generally increasing star formation rate since $\sim$10 Gyr \citep{Bertelli:2001yo}, perhaps with a peak of activity around 3 Gyr \citep{Cignoni:2006om, 2013MNRAS.434.1549R}, though these results are model-dependent and are debated \citep{2009MNRAS.397.1286A}. RC models \citep[e.g.][]{2002MNRAS.337..332S} and observations \citep[e.g.][]{Chen:2017gf} both indicate that RC stars become fainter with increasing age, with a K$_s$ difference of $\sim$0.1 mags between a 3 Gyr and 10 Gyr population. Thus, while our RC model is reasonable, this remains a source of systematic error potentially at the $\sim$0.1 mag level.

We estimate the systematic uncertainty of the Parsec evolution models themselves by comparing the 10 Gyr Parsec RC model a 10 Gyr MIST isochrone RC model, which is built on the MESA stellar evolution models \citep{Choi:2016cr}. The MIST RC model is found to be 0.06 mags brighter in F153M and 0.01 mags larger (i.e., more red) in F127M - F153M color. Further, if one weights the MIST isochrone by a standard Kroupa initial mass function \citep{2001MNRAS.322..231K} and calculates the average magnitudes of stars in the RC portion of the F153M vs. F127M - F153M CMD, the RC model becomes an additional 0.04 mags brighter and 0.003 mags redder. So, the total difference between the adopted Parsec RC model and the IMF-weighted MIST RC model is 0.10 mag in F153M and 0.013 mag in F127M - F153M color. This represents the widest range of systematic error in both color and magnitude from uncertainties in the stellar evolution models themselves.

An error in the average metallicity in the RC sample could impact the absolute magnitudes as well. We have adopted solar metallicity for our RC model, consistent with what has been reported for the Bulge near the Galactic Plane \citep{Zoccali:2003dw, Clarkson:2008hw, Gonzalez:2013il}. However, evidence of a bimodal metallicity distribution in the bulge has been found with peaks around [Fe/H]~$\sim$~-0.3 dex and [Fe/H]~$\sim$~0.3 dex \citep{Hill:2011dq, Bensby:2013qt, Schultheis:2017zk}. RC models for at 10 Gyr population at -0.38 dex $<$ [Fe/H] $<$ 0.2 dex show a $\pm$$\sim$0.1 mag variation relative to the solar metallicity model \citep{2002MNRAS.337..332S}. With these sources of error, we adopt a total systematic error of $\pm$0.1 mags on the RC absolute magnitude due to the uncertainties in the RC star model.

\subsubsection{GC Distance}
An additional source of systematic error is the GC distance, which we've adopted to be 7860 $\pm$ 140 pc from \citet{Boehle:2016rt}. However, a recent compilation and analysis of literature GC distance measurements by \citet{de-Grijs:2016gs} recommends a distance of 8300 $\pm$ 200 (statistical) $\pm$ 400 (systematic) pc. If the GC distance is indeed 8300 pc, then the average RC distance prior is underestimated by $\sim$400 pc. In our analysis, this is equivalent to our RC model being too bright by $\sim$0.1 mags. As a conservative estimate, we adopt an additional systematic error of $\pm$0.1 mags on the RC absolute magnitude due to the GC distance uncertainty.

\subsubsection{Total Systematic Error}
We add the individual sources of systematic error in quadrature for a total systematic error of $\pm$0.14 mags on the RC star absolute magnitude. To assess the impact of this systematic, we change the RC model F153M magnitude by 0.14 mags relative to the model used in $\mathsection$\ref{sec:Wd1Archesresults} and redo the extinction law analysis. We adopt the difference between the extinction law parameters in the new analysis and the original analysis as an estimate of the systematic error and report them in Table \ref{tab:results}. The systematic errors are are approximately equal to or less than the 1$\sigma$ statistical errors.

\subsection{Other Applications and Future Work}
\label{sec:future}
The Wd1+RC law can be applied to stellar populations that similar foreground dust as the Wd1 and Arches field RC stars, namely the spiral arms of the Galaxy in the Galactic Plane. Examples of future applications include studies of stellar populations near the GC, such as the Quintuplet cluster and Young Nuclear Cluster, and the structure and kinematics of the inner Bulge at low galactic latitudes. To aid future use, the Wd1+RC extinction law in several commonly-used filters is provided in Table \ref{tab:filters}. In addition, a python code to generate the law at any wavelength between 0.8 $\mu$m -- 2.2 $\mu$m is available online (see $\mathsection$\ref{sec:Wd1Archesresults}). The law can also be accessed through the astropy-affiliated package \emph{PopStar}, which will soon be released in a beta test.

While a detailed analysis of the dust properties is beyond the scope of this study, the measurement of significant non-power law behavior in the OIR extinction law suggests that there are subtle features that can be used to constrain dust models. The steepness of the law presents a challenge as well, as the classic silicate+graphite grain models \citep[e.g.][]{Mathis:1977qc, Weingartner:2001ul} struggle to produce NIR laws steeper than the canonical C89 law \citep[][]{Moore:2005sy, Fritz:2011cr}. Models that incorporate composite grains with organic refractory material and water ice \cite[e.g.][]{Zubko:2004df} and porous dust structures \cite[e.g.][]{Voshchinnikov:2017lr} are promising in this regard.

\begin{deluxetable}{l c c c}
\tablewidth{0pt}
\tablecaption{Wd1+RC Extinction Law in Different Filters}
\tabletypesize{\scriptsize}
\tablehead{
\colhead{Filter} & \colhead{$\lambda_{pivot}$ ($\mu$m)\tablenotemark{a}} & \colhead{A$_{\lambda}$ / A$_{Ks}$} & \colhead{Filter Ref}\tablenotemark{b}
}
\startdata
2MASS J & 1.239 & 3.69 & 1\\
2MASS H & 1.648 & 1.99 &  1 \\
2MASS K$_s$ & 2.189 & 0.95 &  1 \\
NIRC2 J  & 1.245  & 3.66 & 4  \\
NIRC2 H  & 1.618 & 2.09 & 4  \\
NIRC2 K$_s$ & 2.130 & 1.01 & 4 \\
PS1 i & 0.752 & 11.65 & 2 \\
PS1 z & 0.866 & 8.33 & 2 \\
PS1 y & 0.962 & 6.41 & 2 \\
VISTA Z & 0.880 & 8.00 & 3 \\
VISTA Y & 1.022  & 5.53 & 3 \\
VISTA J & 1.253 & 3.61 & 3 \\
VISTA H & 1.644 & 2.01 & 3 \\
VISTA K$_s$ & 2.145 & 0.99 & 3 \\
\enddata
\label{tab:filters}
\tablenotetext{a}{As defined by \citep{Tokunaga:2005if}}
\tablenotetext{b}{1: \citet{Cohen:2003ay}, 2: \citet{Tonry:2012oz}, 3: \citet{Saito:2012rq}, 4: https://www2.keck.hawaii.edu/inst/nirc2/filters.html}
\end{deluxetable}

A major caveat of this analysis is that it relies on the assumption that the extinction law is the same for Wd1 and the RC stars. While this is supported by our analysis and the literature, future studies are needed to confirm this assumption, especially at shorter wavelengths (e.g. A$_{F814W}$~/~A$_{Ks}$) where R$_V$-like variations might begin to have an effect. Currently, the law normalization is set almost entirely by the RC stars, due to the relatively large uncertainty in the Wd1 cluster distance. A reanalysis of the Wd1 data when the distance is known to higher precision would allow for a constraint on the normalization from the Wd1 stars directly. Similarly, the shape of the extinction law is dominated by the Wd1 data, since the RC observations are limited to just the F127M and F153M filters. Additional observations of the RC stars in filters across a larger wavelength range would confirm that the shape of the law toward the RC stars.

In $\mathsection$\ref{sec:sysError} we show that the systematic errors, mainly coming from uncertainties in the intrinsic RC star properties and GC distance, are roughly the size of the presented error bars in the Wd1+RC extinction law. As a result, a more precise measurement of the extinction law (at least those that use RC stars) is not possible until these uncertainties are addressed. In particular, observational calibrations of RC star models at older ages ($\sim$10 Gyr) are needed to correctly represent the Bulge population, using multiple filters to test the stellar evolution and atmosphere models. Continued progress in understanding the Bulge age/metallicity distribution as well as the GC distance will also decrease the systematics in the extinction law analysis.

\section{Conclusions}
We use \emph{HST} and \emph{VISTA} photometry to measure the OIR (0.8 $\mu$m -- 2.2 $\mu$m) extinction law toward two highly reddened stellar populations: Wd1 and RC stars in the Arches cluster field. The Wd1 sample contains 453 proper-motion selected main sequence stars, a sample 4x larger than previous studies of the extinction law in the cluster. The RC sample contains 813 stars identified in the Arches field CMD. We combine these data sets using a forward modeling Bayesian analysis that simultaneously fits the extinction law and distance of both populations while allowing for systematic offsets in the photometric zeropoints. By combining the samples we measure both the shape and normalization of the extinction law, without making any assumptions regarding its function form.

The best-fit Wd1+RC extinction law is well constrained with typical uncertainties of $\sim$5\% on A$_{\lambda}$~/~A$_{Ks}$. The law is able to reproduce the observed colors of the Wd1 MS stars, where previous extinction laws produce colors that are typically off by 10\% -- 30\%. Contrary to what is often assumed for the OIR, the Wd1+RC law is statistically inconsistent with a single power law, even when only the NIR filters are considered. It is generally steeper (i.e. has larger A$_{\lambda}$~/~A$_{Ks}$ values) than many extinction laws in the literature, being most similar to the \citet{Fitzpatrick:2009ys} law with $\alpha$ = 2.5 and the \citet{Schlafly:2016cr} law with A$_H$ / A$_{Ks}$ = 2.0. Notably, A$_{F125W}$~/~A$_{Ks}$ and A$_{F814W}$~/~A$_{Ks}$ are 18\% and 24\% larger, respectively, than the \citet{Nishiyama:2009fc} law often adopted for the GC/Galactic Bulge, and 4\% and 16\% larger than the previous extinction law derived for Wd1 by \citet{Damineli:2016no}. The new law produces A$_{H}$~/~A$_{Ks}$ = 1.936 $\pm$ 0.08, which is 10\% higher than has been previously measured for RC stars at the GC but is consistent within uncertainties.

Throughout the extinction law analysis we assume an age of 5 Myr for Wd1. We show that varying the cluster age only impacts the cluster distance in the Wd1+RC extinction law fit. We calculate an independent distance to Wd1 using published observations of the eclipsing binary W13 and the new extinction law. The resulting distance of 3937 $\pm$ 332 pc favors an older cluster age of 6 Myr -- 7 Myr, deviating from the cluster distances in the 4 Myr and 5 Myr extinction law models by 3.7$\sigma$ and 2.5$\sigma$, respectively. A detailed analysis of the age and distance of Wd1 is left to a future paper.

This analysis probes the OIR extinction law toward Wd1 ($\ell$ = -20.451$^{\circ}$, $b$ = -0.404$^{\circ}$) and the Arches cluster ($\ell$ = 0.121$^{\circ}$, $b$ = 0.017$^{\circ}$). By necessity, we have assumed that the law is the same for both LOS in this wavelength range, which is supported by our analysis and the literature. Physically, this assumption asserts that the dust causing the extinction for these population have similar properties (in this case, material from foreground spiral arms in the Galactic plane). While future studies of Wd1 and the RC stars are required to verify this assumption, the Wd1+RC law is the best available for highly reddened stellar populations with similar foreground material, such as the Quintuplet cluster and Young Nuclear Cluster. For ease of use, the extinction law in several commonly-used filter sets is provided in Table \ref{tab:filters} and a python code to generate the law is available online (see $\mathsection$ \ref{sec:Wd1Archesresults}).

We demonstrate that the methodology developed in this paper allows for a highly detailed measurement of the extinction law. Such measurements will become critical in light of upcoming space-based infrared observatories such as the James Web Space Telescope (JWST) and Wide Field Infrared Survey Telescope (WFIRST), which will push infrared observations into increasingly extinguished regions with high precision.

\acknowledgements
The authors thanks the anonymous referee for their helpful comments that greatly improved the clarity of the paper. M.W.H. and J.R.L. acknowledge support from NSF AAG (AST-1518273) and HST GO-13809. This work is based on observations made with the NASA/ESA Hubble Space Telescope, obtained at the Space Telescope Science Institute, which is operated by the Association of Universities for Research in Astronomy, Inc., under NASA contract NAS 5-26555. The Arches field observations observations are associated with programs 11671, 12318, and 12667, and the Westerlund 1 observations are associated with programs 10172, 11708, and 13044. It also uses data products from observations made with ESO Telescopes at the La Silla or Paranal Observatories under ESO programme ID 179.B-2002. This research has made extensive use of the NASA Astrophysical Data System.

\facilities{HST (WFC3-IR, ACS-WFC), ESO:VISTA}
\software{AIROPA \citep{Witzel:2016lq}, AstroPy \citep{Astropy-Collaboration:2013kx}, DOLPHOT \citep{Dolphin:2000mw}, extcurve\_s16.py (http://faun.rc.fas.harvard.edu\\/eschlafly/apored/extcurve\_s16.py), Matplotlib \citep{Hunter:2007}, SciPy \citep{SciPy}, Starfinder \citep{Diolaiti:2000rc}, Tiny Tim \citep{Krist:2011ev}}

\clearpage

\appendix

\section{Extinction Law Definitions}
\label{app:defs}
In this appendix we describe the extinction law terms used in the paper. The parameter A$_{\lambda}$ is the amount of extinction (in magnitudes) observed toward a source at wavelength $\lambda$:

\begin{equation}
m_{obs} = M_0 + \mu + A_{\lambda}
\end{equation}

where $m_{obs}$ and $M_0$ are the observed and absolute magnitudes of a source, respectively, and $\mu$ is the distance modulus. We present the extinction law as the ratio A$_{\lambda}$/A$_{K_s}$, which can be generally written as:

\begin{equation}
A_{\lambda} / A_{K_s} = b f(\lambda) + c
\end{equation}

where $f(\lambda)$ is the wavelength-dependent shape of the extinction law and $b$ and $c$ together comprise the normalization factors. The shape of the extinction law is often constrained via stellar color-excess ratios, where the normalization factors conveniently cancel out:

\begin{equation}
\label{eq:ce}
\frac{E(\lambda_1 - \lambda_2)}{E(\lambda_2 - \lambda_3)} = \frac{A_{\lambda_1} - A_{\lambda_2}}{A_{\lambda_2} - A_{\lambda_3}} = \frac{f(\lambda_1) - f(\lambda_2)}{f(\lambda_2) - f(\lambda_3)}
\end{equation}

While being directly observed quantities, color excess ratios are not continuous as a function of wavelength and depend on the filters used. This makes comparing color-excess ratios across different studies difficult. We present an alternative approach based on the derivative of the extinction law with respect to 1/$\lambda$:

\begin{equation}
\nabla_{1/\lambda} = \frac{\partial(A_{\lambda} / A_{Ks})}{\partial(1/\lambda)} = -b*\lambda^2*f'(\lambda)
\end{equation}

where $f'(\lambda) = \frac{\partial(f(\lambda))}{\partial(\lambda)}$. We define the parameter $S_{1/\lambda}$ such that it only depends on the extinction law shape and the ratio of $\lambda$ to 2.14 $\mu$m, which we use as a reference wavelength:

\begin{equation}
\label{eq:Slambda}
S_{1/\lambda} = \frac{\nabla_{1/\lambda}}{\nabla_{1/2.14\mu m}}  = \frac{f'(\lambda)}{f'(2.14 \mu m)} * \left(\frac{\lambda}{2.14}\right)^2
\end{equation}

To see the relationship between $S_{1/\lambda}$ and the color excess ratio, we evaluate Equation \ref{eq:ce} in the limit where $\lambda_2$~-~$\lambda_1$~=~$\Delta\lambda_1$ and  $\lambda_3$~-~$\lambda_2$~=~$\Delta\lambda_2$ are small and set $\lambda_2$ = 2.14 $\mu$m:

\begin{equation}
\frac{E(\lambda_1 - \lambda_2)}{E(\lambda_2 - \lambda_3)} = \frac{f(\lambda_1) - f(\lambda_1 + \Delta \lambda_1)}{f(\lambda_2) - f(\lambda_2 + \Delta \lambda_2)} = \frac{f'(\lambda_1)}{f'(\lambda_2)} \frac{\Delta \lambda_1}{\Delta \lambda_2} = S_{1/\lambda}\left(\frac{\lambda_2}{\lambda_1}\right)^2  \frac{\Delta \lambda_1}{\Delta \lambda_2}
\end{equation}

\section{Simulated Data Tests}
\label{app:art}
We simulate a set of Wd1 MS and Arches RC star observations with known extinction properties in order to test the extinction law analysis. The synthetic photometry is constructed from stellar models as described in $\mathsection$\ref{sec:Iso}. A total of 400 stars are randomly drawn from this isochrone (the approximate size of the Wd1 sample) and extinction is applied using a \citet{Nishiyama:2009fc} extinction law. To replicate differential extinction, the total extinction for each star is drawn from a Gaussian distribution centered at A$_{Ks}$ = 0.7 mags with a width of dA$_{Ks}$ = 0.15 mags. This distribution was found to broadly reproduce the spread of the real observations in CMD and 2CD space. Finally, realistic photometric errors are applied to the simulated measurements based on the median photometric error calculated as a function of magnitude for the observations in each filter. The photometry of each star is perturbed by a value drawn from a Gaussian distribution with $\mu$ = 0 and standard deviation equal to the appropriate photometric error at that star's magnitude in the given filter (typically $\sim$0.01 mag).

The synthetic RC stars are generated in a similar manner, adopting the RC stellar model from $\mathsection$\ref{sec:Iso} and creating a sample of 900 stars with extinction values uniformly distributed between 2.7 mag $<$ A$_{Ks}$ $<$ 3.5 mag using a \citet{Nishiyama:2009fc} law. This dA$_{Ks}$ was found to reproduce the color range of the observed RC sequence in CMD space. Initially, the stars are generated at the same distance, namely the d$_{rc}$ value defined by the user. Then, the photometry of each star is perturbed by a value drawn from a Gaussian distribution with a width equal $\sigma_{rc}$, also defined by the user. This perturbation, which is the same in both filters, represents the impact of the unknown distance of the particular star as discussed in $\mathsection$\ref{sec:RCsamp}. Like Wd1, realistic photometric errors are also added to the photometry. The RC sample is then restricted to stars that fall within $\pm$0.3 mag of the reddening vector in both of CMDs, matching the criteria adopted for the observed RC sample.

\begin{figure*}
\begin{center}
\includegraphics[scale=0.4]{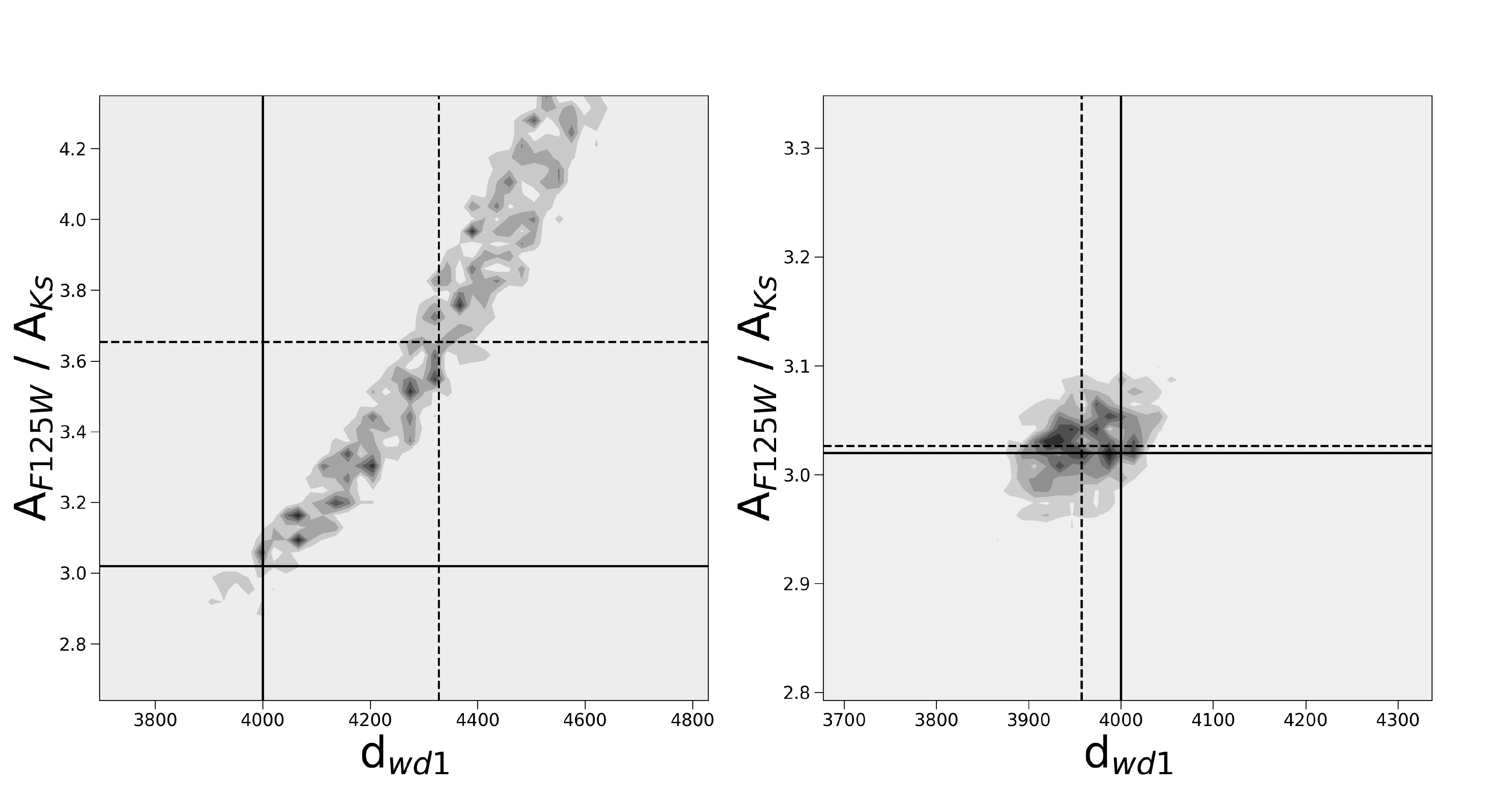}
\end{center}
\caption{The marginalized two-dimensional posterior probability distributions for A$_{F125W}$~/~A$_{Ks}$ vs. d$_{wd1}$ in the simulated cluster tests. These posteriors show the degeneracy between the extinction law and cluster parameters (d$_{wd1}$ in this case). The solid black lines denote the input value for the simulated cluster while the dotted black lines denote the best-fit value from the fit. \emph{Left:} The posterior for the Wd1-only simulated analysis, which has a large degeneracy due to the uncertainty in the cluster distance and thus normalization of the law. \emph{Right:} The posterior for the Wd1+RC simulated analysis, where the extinction law normalization is constrained by the RC stars and thus the degeneracy is broken.
\label{fig:art_new}
}
\end{figure*}

First, we test the extinction law fitter using the simulated Wd1 MS sample only; that is, $\mathcal{L}_{tot} =  \mathcal{L}_{wd1}$ in equation \ref{eq:bayes}. The synthetic star catalog is subjected to the same set of cuts as the observed catalog discussed in $\mathsection$\ref{sec:Wd1samp}. We adopt the same priors as is used for the real data except for the RC parameters which aren't used in the model. The resulting posterior distributions show a large degeneracy as a wide range of extinction laws are allowed for different combinations of A$_{Ks}$ and d$_{wd1}$ (Figure \ref{fig:art_new}). Generally, the synthetic observations can be reproduced by a more distant cluster model with a lower A$_{Ks}$ and steeper extinction law, or a closer cluster with a higher A$_{Ks}$ and shallower law. This degeneracy caused by the uncertainty in the normalization.

Next, we test the fitter using only the simulated RC sample, such that $\mathcal{L}_{tot} =  \mathcal{L}_{RC}$ in equation \ref{eq:bayes}. We similarly subject the stars to the same cuts as the observed sample and adopt the same priors. None of the Wd1 cluster parameters are used, and the extinction law itself is limited to A$_{F125W}$ / A$_{F160W}$, where the RC data has constraining power. The resulting posterior distributions show that all input parameters are recovered to within 1$\sigma$.

Finally, we test the fitter using both the simulated Wd1 MS and Arches RC data sets. The advantage of the RC sample is that the distance distribution is known to much higher precision than that of Wd1, allowing for a tighter constraint on the normalization of the extinction law (Figure \ref{fig:art_new}). The RC sample distribution is anchored to the GC, which has a distance that is known to within $\sim$2\% \citep{Boehle:2016rt}, compared to the Wd1 distance uncertainty of $\sim$18\%. The fit recovers the input extinction law parameters to within 1$\sigma$, with an uncertainty in A$_{\lambda}$ / A$_{Ks}$ ranging from 0.14 at F814W to 0.02 at F160W. The remaining parameters in the model (A$_{Ks}$, d$_{wd1}$, d$_{rc}$, $\sigma_{rc}$, and the zeropoint offsets) are also recovered within 1$\sigma$. The set of simulated cluster results is provided in Table \ref{tab:art}.

\begin{deluxetable}{l c c c c}
\tablewidth{0pt}
\tablecaption{Simulated Data Results}
\tabletypesize{\scriptsize}
\tablehead{
\colhead{Parameter} & \colhead{Input} & \colhead{Prior\tablenotemark{a}} & \colhead{Wd1-only} &
\colhead{Wd1+RC}
}
\startdata
A$_{F814W}$ / A$_{Ks}$ & 8.87 & U(4, 14) & 11.31 $\pm$ 1.82 & 8.86 $\pm$ 0.12\\
A$_{y}$ / A$_{Ks}$ & 6.00 & U(4, 14) & 7.34 $\pm$ 1.13 &  5.98 $\pm$ 0.09 \\
A$_{F125W}$ / A$_{Ks}$ & 3.02 & U(1, 6) & 3.65 $\pm$ 0.47 & 3.03 $\pm$ 0.03 \\
A$_{F160W}$ / A$_{Ks}$ & 1.93 & U(1, 6) & 2.22 $\pm$ 0.21 & 1.93 $\pm$ 0.02 \\
A$_{Ks}$ & 0.70 & U(0.3, 1.4) & 0.53 $\pm$ 0.10 & 0.71 $\pm$ 0.02 \\
A$_{[3.6]}$ / A$_{Ks}$ & 0.50 & G(0.50, 0.05) & 0.50 $\pm$ 0.03 & 0.50 $\pm$ 0.04 \\
d$_{wd1}$ & 4000 & G(4000, 700)  & 4328 $\pm$ 193 &  3958 $\pm$ 45 \\
d$_{rc}$ & 8000 & G(8000, 160)  & -- & 8009 $\pm$ 105 \\
$\sigma_{rc}$& 0.2 & G(0.2, 0.01)  & -- & 0.2 $\pm$ 0.006 \\
ZP$_{Ks}$  & 0  & G(0, 5x10$^{-3}$) & -8x10$^{-4}$ $\pm$ 4x10$^{-3}$ & -1.3x10$^{-3}$ $\pm$ 3x10$^{-3}$ \\
ZP$_{F160W}$ & 0  & G(0, 8.8x10$^{-4}$) & 0.0 $\pm$ 1x10$^{-3}$ & -1x10$^{-4}$ $\pm$ 1x10$^{-3}$ \\
ZP$_{F153M}$ & 0  & G(0, 7.2x10$^{-4}$) & -- &  0.0 $\pm$ 1x10$^{-4}$ \\
ZP$_{F127M}$ & 0  & G(0, 7.1x10$^{-4}$) & -- &  0.0 $\pm$ 1x10$^{-4}$ \\
ZP$_{F125W}$ & 0  & G(0, 1.2x10$^{-3}$) & 1x10$^{-4}$ $\pm$ 1x10$^{-3}$ & 0.0 $\pm$ 1x10$^{-3}$ \\
ZP$_{F1814W}$ & 0  & G(0, 5.2x10$^{-4}$) & 0.0 $\pm$ 1x10$^{-4}$ & 0.0 $\pm$ 1x10$^{-4}$ \\
\enddata
\label{tab:art}
\tablenotetext{a}{Uniform distributions: U(min, max), where min and max are bounds of the distribution; Gaussian distributions: G($\mu$, $\sigma$), where $\mu$ is the mean and $\sigma$ is the standard deviation}
\end{deluxetable}

\section{Extinction Law Fit Posteriors}
\label{app:posteriors}
\begin{figure*}
\begin{center}
\includegraphics[scale=0.4]{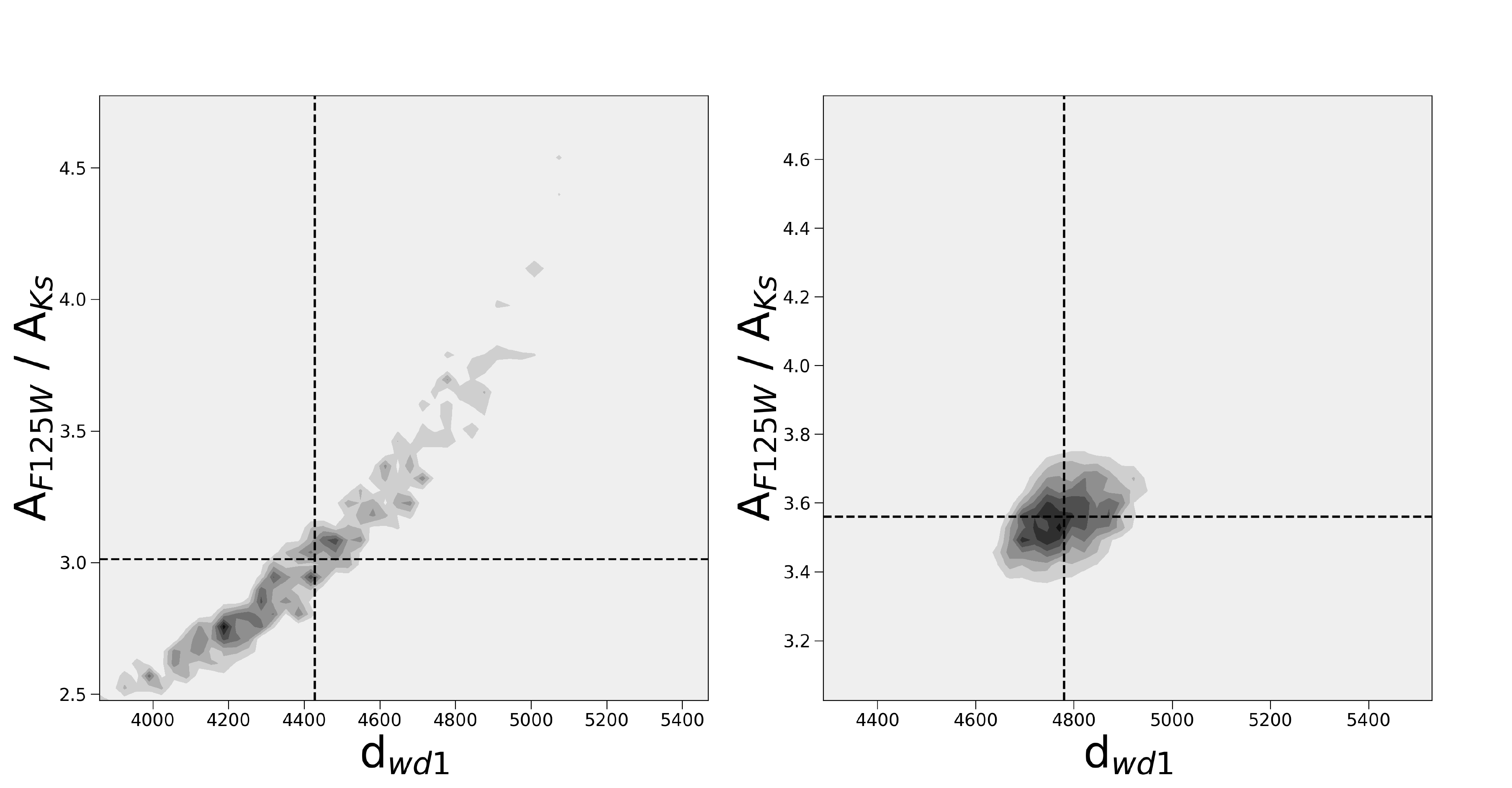}
\end{center}
\caption{
The marginalized two-dimensional posterior probability distributions  for A$_{F125W}$~/~A$_{Ks}$ vs. d$_{wd1}$ in the extinction law analysis. The dotted black lines denote the best-fit value from the fit. \emph{Left:} The posterior for the Wd1-only analysis, which suffers from a large degeneracy due to the uncertainty in the extinction law normalization. \emph{Right:} The posterior for the Wd1+RC analysis, where the law normalization has been constrained by the RC stars. The posteriors for the real cluster analyses are similar to what we expect based on the simulated cluster analyses.
\label{fig:wd1_deg}
}
\end{figure*}

In this appendix we present a representative example of the posterior distributions for the extinction law fits, specifically the A$_{F125W}$~/~A$_{Ks}$ two-dimensional posterior distributions. Figure \ref{fig:wd1_deg} shows the posteriors for the Wd1-only fit  ($\mathsection$\ref{sec:Wd1results}) and the Wd1 + RC fit ($\mathsection$\ref{sec:Wd1Archesresults}).

\end{document}